\documentclass[12pt]{iopart}

\expandafter\let\csname equation*\endcsname\relax
\expandafter\let\csname endequation*\endcsname\relax

\usepackage{graphicx}
\usepackage{mathrsfs}
\usepackage{graphics}
\usepackage{amsmath}
\usepackage{color}
\usepackage[format=plain,labelsep=period]{caption}
\begin{document}

\title{Dephasing effect promotes the appearance of quantized Hall plateaus}

\author{Jing-Yun Fang$^{1,2}$, Ai-Min Guo$^{3}$ and Qing-Feng Sun$^{1,2,4,*}$}

\address{$^1$ International Center for Quantum Materials, School of Physics, Peking University, Beijing 100871, China}
\address{$^2$ CAS Center for Excellence in Topological Quantum Computation, University of Chinese Academy of Sciences, Beijing 100190, China}
\address{$^3$ Hunan Key Laboratory for Super-microstructure and Ultrafast Process, School of Physics and Electronics, Central South University, Changsha 410083, China}
\address{$^4$ Beijing Academy of Quantum Information Sciences, West Bld.\#3,No.10 Xibeiwang East Rd., Haidian District, Beijing 100193, China}
\ead{sunqf@pku.edu.cn}
\vspace{10pt}
\begin{indented}
\item[]{\today}
\end{indented}

\begin{abstract}
The quantum Hall effect (QHE) is a topologically protected phenomenon which has been observed in various systems.
In experiments, the size of Hall bar device to realize the QHE is generally much larger than
the phase coherence length, in which the quantum coherence of electrons is destroyed. Here, we theoretically study the influence of dephasing effect on the quantized Hall (QH) plateaus. We find that the QH plateau disappears in perfectly quantum coherent systems if the coupling between leads and central region is imperfect. The Hall resistance is very large and strongly oscillates instead of presenting the QH plateau in this case. However, by introducing the dephasing, the Hall resistance decreases and the QH plateau appears gradually. Similar results can also be observed for the quantum anomalous Hall effect. Our results propose that
dephasing effect promotes the appearance of QH plateaus, which opens a new topic of the dephasing effect on topological systems.
\end{abstract}


\section{Introduction}
Topology is a fundamentally important concept in condensed matter physics,
which asserts backscattering-immune edge (or surface)
states~\cite{RevModPhys.81.109,RevModPhys.82.3045}.
Quantum Hall effect (QHE), a prime topologically protected phenomenon
which is first discovered in a two-dimensional electron gas (2DEG) system~\cite{PhysRevLett.45.494}, has attracted great attention during
the past decades~\cite{PhysRevLett.48.1559,PhysRevLett.50.1395,RevModPhys.58.519,Prange,Nature.569.537,Nature.579.56}. In the presence of a strong perpendicular magnetic field, the spectrum evolves into a series of impurity-broadened discrete Landau levels,
with the extended states close to Landau levels while the
localized far from Landau levels~\cite{PhysRevB.23.5632,PhysRevB.25.2185}. When Fermi level is in the localized states, a quantized Hall (QH) plateau is developed,
while the Fermi level is in the extended states between two adjacent QH plateaus.
The plateau-to-plateau transition is therefore a localization-delocalization transition,
and suitable dephasing can broaden the plateau transition region~\cite{RevModPhys.67.357,RevModPhys.69.315,PhysRevLett.61.1297,PhysRevLett.67.883}.
This QH plateau is characterized by Chern number~\cite{PhysRevLett.49.405},
a nonzero integer that counts the number of chiral edge states. Since the QH resistance is determined by the topological properties of a system,
it is insensitive to the details of samples.
In particular, the QH resistance is so accurate that it can be used as a resistance standard~\cite{Rep.Prog.Phys.64.1603}
and determine the fine-structure constant~\cite{PhysRevLett.45.494},
making it plays a crucial role in metrology.

Graphene, a monolayer hexagonal lattice of carbon atoms,
has been extensively investigated since it is successfully extracted
from graphite~\cite{Science.306.666,RevModPhys.83.407,Nature.583.375,Nat.Mater.6.183}. Graphene has a unique band structure in which the conduction and
valence band intersect at two nonequivalent points $K$ and $K'$,
the corners of the hexagonal first Brillouin zone.
Around these two points (also called ``Dirac points"),
graphene has a linear dispersion relation. Its quasiparticles obey
the massless Dirac-type equation where the speed of light is replaced
by the Fermi velocity of graphene ($v_{F}\approx 10^6$ m/s)~\cite{PhysRevB.73.235411}. Such a unique band structure leads
to many extraordinary properties~\cite{Nature.438.197,Nature.438.201,RevModPhys.80.1337,PhysRevLett.97.187401,NatPhys.2.620}.
One of the most remarkable properties is the QHE~\cite{Nature.438.197,Nature.438.201,PhysRevLett.95.146801}.
Since the linear dispersion relation in graphene has electron-hole symmetry,
the zeroth Landau level at Dirac points is electron-hole degenerate under a strong magnetic field. This gives rise to a special QHE in graphene,
in which the Hall conductance has a half-integer values of $g(n+1/2)e^2/h$, where $g=4$ is the spin and valley degeneracy and $n$ is the index of Landau levels~\cite{Nature.438.197,Nature.438.201}. This QHE in graphene can be realized even at room temperature~\cite{Science.315.1379}. Besides, graphene has excellent mechanical strength~\cite{Science.321.385}, nanoscale scalability~\cite{NatNanotechnol.3.270}, superior thermal conductivity~\cite{NanoLett.8.902}, high carrier mobility~\cite{SolidStateCommun.146.351,NatNanotechnol.3.491}
and electrical conductivity~\cite{Science.332.1537,Science.335.1326}, etc., making it attractive for profound applications in future nanoelectronics.

Experimentally,
the device size (denoted by $N$ hereafter) to realize QHE
generally ranges from tens to hundreds of
microns~\cite{Nature.438.197,Nature.438.201,Nature.575.628}.
While the phase coherence length $L_{\varphi}$ is usually about several microns or hundreds of nanometers at the low temperature~\cite{PhysRevB.78.125409,Science.312.1191}.
Obviously, the device size is much longer than
the phase coherence length ($N\gg L_{\varphi}$),
thus electrons can't keep perfect quantum coherence.
In addition, realization of the QHE requires the upper and lower chiral edge states separate in space, i.e., the device size needs to be
much larger than the magnetic length $l_{B}$.
So a strong magnetic field $B$ is needed, e.g., $B\approx5$T, which corresponds to a magnetic length $l_{B}\approx11$ nm ($N\gg l_{B}$).
Under the above two conditions ($N \gg L_{\varphi} \gg l_{B}$), the QHE can be observed in experiments, although electrons can't keep perfect quantum coherence. By intuitive, this leads to two speculations. 
(i) Since the QHE can be observed when electrons can't keep perfect quantum coherence, 
it should be better observed if perfect quantum coherence 
is guaranteed (i.e., $L_{\varphi}\gg N\gg l_{B}$).
(ii) In real devices, dephasing effect inevitably exists because of electron-phonon interaction, electron-electron interaction, etc.~\cite{PhysRevB.78.045322,PhysRevA.41.3436}. 
It will induce the phase-relaxation and randomize the phase of incident electrons~\cite{PhysRevB.75.081301}, thereby destroying the quantum coherence of electrons. 
Previous studies on the scaling in plateau-to-plateau transition in QHE systems have found that dephasing can broaden the plateau transition region and thus shorten the QH plateau~\cite{RevModPhys.67.357,PhysRevLett.61.1297}.
However, since the QH plateaus can be observed for $N \gg L_{\varphi}$ in real devices with imperfect lead-central region coupling, it seems that dephasing has no bad influence on the appearance of QH plateaus. 
Of course, the validity of these two speculations remains to be further studied. 
Recently, Marguerite \textit{et al.} experimentally found the QHE accompanied
with dissipation along the chiral edge states in graphene,
which violates the principle that QHE is dissipationless~\cite{Nature.575.628}.
Therefore, the speculations that QHE can better appear in perfectly
quantum coherent systems and dephasing effect has no bad influence on the appearance of QH plateaus should be considered as well.

\begin{figure}
  \includegraphics[width=0.8\textwidth]{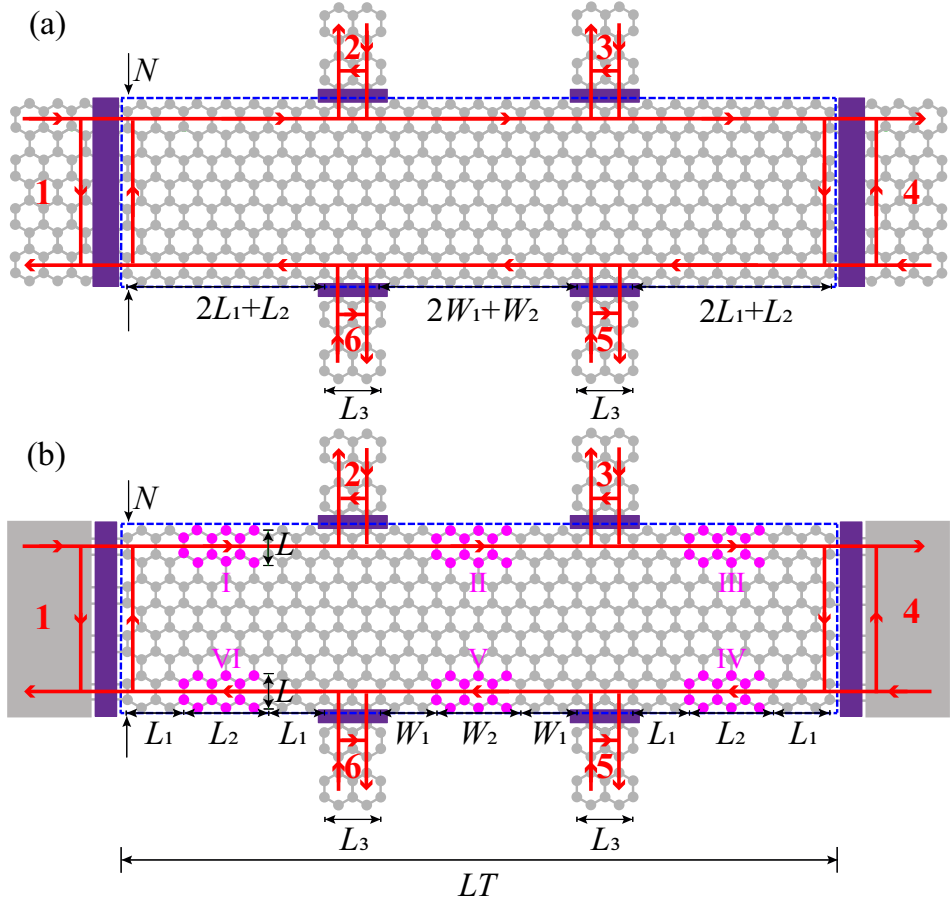}
  \centering
  \caption{Illustration of a six-terminal graphene Hall bar device. Leads 2, 3, 5, 6 are
  semi-infinite graphene nanoribbons, and the box with blue dotted lines
  is the central region. The red solid lines with arrows represent the propagation
  of electrons along the edge states.
  The purple areas represent the interfacial resistances
  when the lead-central region coupling is imperfect. In these two diagrams, the device sizes are $N=16$, $L_1=W_1=2$, $L_2=W_2=3$, $L_3=2$, and $L=4$. Leads 1 and 4 are semi-infinite graphene nanoribbons in (a), but metal leads in (b). And the magenta-sites in (b) indicate that virtual leads are coupled to these sites.}
  \label {fig:1}
  \end{figure}

In this paper, we solve these two aforementioned speculations.
By using the tight-binding model, combining Landauer-B\"{u}ttiker formula with nonequilibrium Green's function method~\cite{PhysRevLett.68.2512,PhysRevB.50.5528,Datta}, we consider six-terminal Hall bar devices and focus on the influence of dephasing effect on the QH plateaus. Our results show that the QH plateaus disappear in perfectly quantum coherent systems with imperfect lead-central region coupling, although the device size is much larger than the magnetic length as in most experiments. Instead of presenting the QH plateau, Hall resistance strongly oscillates with magnetic field or Fermi energy. However, with the increase of dephasing, the QH plateau appears gradually. This indicates that dephasing effect promotes the appearance of QH plateaus. Furthermore, we also study the QHE in 2DEG system and the quantum anomalous Hall effect (QAHE)~\cite{RevModPhys.82.1539,PhysRevLett.101.146802}, and find that
the QH plateau also disappears in perfectly quantum coherent systems
but appears when strong dephasing exists.

The rest of the paper is organized as follows.
In section 2, we describe the model of the graphene Hall bar
and the method we employ to calculate the Hall and longitudinal resistance.
In section 3, we show that QHE appears when leads are perfectly coupled to central region.
In section 4, we show that QHE always disappears in the graphene system with imperfect lead-central region coupling, although perfect quantum coherence is guaranteed.
In sections 5 and 6, we show that dephasing effect promotes the appearance of QHE in graphene systems without and with disorder, respectively. The QHE in 2DEG system and QAHE are studied in sections 7 and 8, respectively.
Finally, a conclusion is given in section 9.

\section{Model and method}

We consider six-terminal devices consisting of a central Hall bar coupled to left and right leads (i.e., leads 1, 4).
To better present our results, here we study two different models for comparison.

(1) Left/right leads and central region are the same material (see figure~\ref{fig:1}(a)).

In this case, the Hamiltonian of a six-terminal device consisting of a graphene Hall bar
coupled to left and right graphene leads can be expressed as
\begin{equation}\label{eq:1}
H=H_{\textrm G}+H_{\textrm M1}+H_{\textrm C1},
\end{equation}
where $H_{\textrm G}$, $H_{\textrm M1}$, and $H_{\textrm C1}$ are, respectively, the Hamiltonian of central Hall bar, left and right leads 1, 4, and coupling between central region and leads 1, 4.
In the tight-binding representation, $H_{\textrm G}$ has the form~\cite{PhysRevLett.104.066805}
\begin{equation}\label{eq:2}
H_{\textrm G}=\sum_{\bf i}(\epsilon_{\bf i}-E_{f})c_{\bf i}^{\dag}c_{\bf i}-\sum_{\langle{\bf ij}\rangle}
t_{g}e^{i\phi_{\bf ij}}c_{\bf i}^{\dag}c_{\bf j},
\end{equation}
where $c_{\bf i}^\dag$ ($c_{\bf i}$) is the creation (annihilation) operator at site ${\bf i}$. $\epsilon_{\bf i}=\epsilon_{0}$ is the energy of Dirac point (i.e., the on-site energy).
Below $\epsilon_{0}$ is set to zero as the energy reference point.
$E_{f}$ is the Fermi energy, which can be tuned by the gate voltage experimentally. The second term in $H_{\textrm G}$ describes
the nearest-neighbor hopping with $t_{g}$ the hopping energy.
Because of the existence of a perpendicular magnetic field $B$,
a phase $\phi_{\bf ij}=\int_{\bf i}^{\bf j}{\vec{A}\cdot{d\vec{l}}/\phi_{0}}$
is added in the hopping element,
with vector potential $\vec{A}=(-By,0,0)$ and $\phi_{0}=\hbar/e$. $H_{\textrm M1}=\sum_{\bf i}(\epsilon_{\bf i}-E_{f})b_{\bf i}^{\dag}b_{\bf i}-\sum_{\langle{\bf ij}\rangle}
t_{g}e^{i\phi_{\bf ij}}b_{\bf i}^{\dag}b_{\bf j}$ is the Hamiltonian of leads 1, 4, with $b_{\bf i}^\dag$ ($b_{\bf i}$) the creation (annihilation) operator. $H_{\textrm C1}=-t_{c1}\sum_{\langle{\bf ij}\rangle}e^{i\phi_{\bf ij}}b_{\bf i}^{\dag}c_{\bf j}+\textrm{H.c.}$ is the coupling between central region and leads 1, 4 with $t_{c1}$ the coupling strength.
Note that here the coupling strength $t_{c1}$ can be equal to
or not equal to $t_g$, which corresponds to the cases of perfect or imperfect coupling.

(2) Left/right leads and central region are different materials (see figure~\ref{fig:1}(b)).

In this case, we also study the influence of dephasing effect on the QH plateaus. Here we simulate dephasing processes
by introducing the B\"{u}ttiker's virtual leads~\cite{PhysRevLett.57.1761,PhysRevB.77.115346,PhysRevLett.103.036803,PhysRevLett.63.1857,PhysRevLett.64.216,PhysRevLett.61.589}.
Then, the Hamiltonian of a six-terminal device consisting of a graphene Hall bar
coupled to left and right metal leads can be expressed as
\begin{equation}\label{eq:3}
H=H_{\textrm G}+H_{\textrm M2}+H_{\textrm C2}+H_{\textrm D2}.
\end{equation}
Here $H_{\textrm G}$ is the Hamiltonian of the central graphene Hall
bar, which is the same as that in equation~(\ref{eq:2}).
$H_{\textrm M2}$, $H_{\textrm C2}$, and $H_{\textrm D2}$ are, respectively, the Hamiltonian of left/right leads,
coupling between graphene and left/right leads,
B\"{u}ttiker's virtual leads and their couplings to
central sites. In experiments, the central Hall bar and leads
are usually different materials~\cite{Science.315.1379,Nature.575.628}.
So we consider the left/right leads are metals with a square-lattice Hamiltonian
$H_{\textrm M2}=\sum_{\bf i}\epsilon_{2}b_{\bf i}^{\dag}b_{\bf i}-\sum_{\langle{\bf ij}\rangle}t_{m2}b_{\bf i}^{\dag}b_{\bf j}$,
where $\epsilon_{2}$ is the on-site energy, and $t_{m2}$ is the hopping energy. The lead-graphene coupling Hamiltonian is
$H_{\textrm C2}=t_{c2}\sum_{\langle{\bf ij}\rangle}b_{\bf i}^{\dag}c_{\bf j}+\textrm{H.c.}$, with $t_{c2}$ the coupling strength.
Since the lattices of the Hall bar and the leads are different types, reflection usually occurs at their interfaces.
The Hamiltonian of virtual leads and their couplings to
central sites
is $H_{\textrm D2}=\sum_{{\bf i},k}\epsilon_{k}a_{{\bf i}k}^{\dag}a_{{\bf i}k}+(t_{k}a_{{\bf i}k}^{\dag}c_{\bf i}+\textrm{H.c.})$,
with $a_{{\bf i}k}^{\dag}$ ($a_{{\bf i}k}$) the creation (annihilation) operator in the virtual leads and $t_k$ the virtual lead-graphene coupling strength. Here we consider that virtual leads only exist in the magenta-site regions (i.e., regions {\uppercase\expandafter{\romannumeral1}}-{\uppercase\expandafter{\romannumeral6}} in figure~\ref{fig:1}(b)), where each site is attached by a virtual lead.

\begin{figure}
  \includegraphics[width=0.8\textwidth]{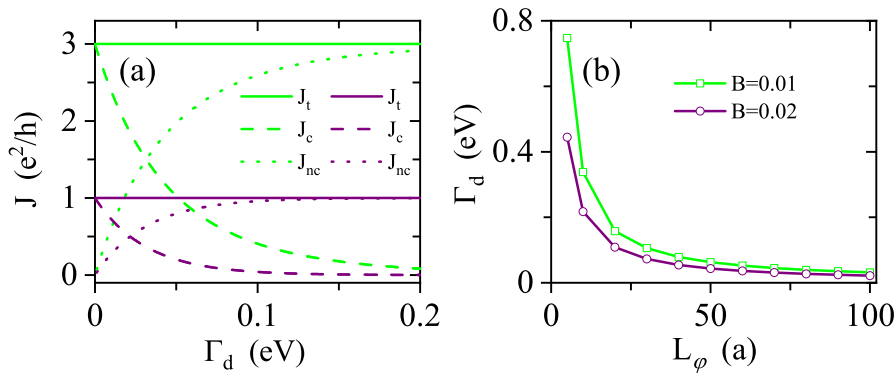}
  \centering
  \caption{ (a) Total current $J_{t}$, coherent component $J_{c}$,
  and incoherent component $J_{nc}$ vs the dephasing strength $\Gamma_{d}$
  in the two-terminal graphene nanoribbon system for the magnetic field $B=0.01$ ($\nu=3$, green lines)
  and $B=0.02$ ($\nu=1$, purple lines). The system sizes are $N=120$, $LT=100$, $L=10$.
  (b) The dephasing strength $\Gamma_{d}$ vs phase coherence length $L_{\varphi}$
  for different magnetic fields.}
  \label{fig:2}
  \end{figure}

The current in lead $p$ (real or virtual) at the low temperature limit
can be obtained from the Landauer-B\"{u}ttiker formula~\cite{Datta,PhysRevLett.95.136602,PhysRevLett.64.220}
\begin{equation}\label{eq:4}	
J_{p}= \frac{e^2}{h}\sum_{q\ne p}T_{p,q}(V_{p}-V_{q}),
\end{equation}
where $V_{p}$ is the voltage in lead $p$.
$T_{p,q}=\textmd{Tr}[{\bf \Gamma}_{p}{\bf G}^r{\bf \Gamma}_{q}{\bf G}^a]$ is the transmission coefficient from lead $q$ to lead $p$,
where the Green's function ${\bf G}^r(E)=[{\bf G}^a(E)]^\dag=[E {\bf I}-{\bf  H}_{\rm cen}-\sum_{p}{\bf \Sigma}_{p}^r]^{-1}$
and the linewidth function
${\bf \Gamma}_{p}(E)=i[{\bf \Sigma}_{p}^r(E)-{\bf \Sigma}_{p}^{a}(E)]$.
${\bf H}_{\rm cen}$ is the Hamiltonian of the central region.
${\bf \Sigma}_{p}^r(E) =[{\bf \Sigma}_{p}^{a}(E)]^\dagger$ is the retarded self-energy
due to the coupling to lead $p$. For real lead $p$, the self-energy ${\bf \Sigma}_{p}^r$ can be calculated
numerically.
For virtual lead $p$, ${\bf \Sigma}_{p}^r=-\frac{i}{2} \Gamma_{d}$,
with the dephasing strength $\Gamma_{d}=2\pi\rho t_k^2$
and $\rho$ the density of states in the virtual leads~\cite{PhysRevLett.57.1761,PhysRevB.77.115346}.
In the wide-band approximation, $\Gamma_{d}$ is independent of energy. When a small bias is applied between leads 1 and 4 with $V_1=-V_4=V$,
the current flows along the longitudinal direction.
The leads 2, 3, 5, 6 are set as voltage probes with zero net currents.
The currents in the virtual leads are also zero, 
because electrons only lose phase memories when going 
into and coming back from the virtual leads~\cite{PhysRevLett.57.1761}. Since the energy of the electron entering the virtual lead and 
the energy of the electron coming out from the virtual lead 
are likely to be unequal, this is an inelastic scattering process
and the dephasing occurs.
Combining equation~(\ref{eq:4}) with all these boundary conditions,
the voltage $V_{p}$ and the longitudinal current ($J_{1}=-J_{4}$) can be obtained self-consistently.
Then the Hall resistance $R_{H}=(V_{2}-V_{6})/J_{1}$
and the longitudinal resistance $R_{xx}=(V_{2}-V_{3})/J_{1}$
can be calculated straightforwardly.

In the numerical calculations, we set the model parameters $t_{g}=2.75$, $\epsilon_{2}=3$, and $t_{m2}=1$, with the unit being eV. The zigzag edge graphene ribbon
is considered and the results still hold for the armchair one.
The magnetic field is expressed by
$2\phi=(3\sqrt{3}/2)a^2B/\phi_{0}$, with $(3\sqrt{3}/2)a^2B$ being the magnetic flux threading a single
hexagon.
When $B=0.01$, $l_{B}\approx11$.
The device sizes are $L_1=W_1=40$, $L_2=W_2=120$, $L_3=2$, and $L=10$. Thus $(N,LT)=(120,604)$ is much larger than $l_{B}$ for $B>0.001$, in which the Landau levels
well form and the chiral edge states appear.
Notice that lead 2 (or 3, 5, 6) laterally couples with central region,
so reflection usually occurs at the interface.

The dephasing strength $\Gamma_{d}$ can be directly related to the phase coherence length $L_\varphi$~\cite{PhysRevB.77.115346,PhysRevLett.103.036803},
an observable parameter in experiment.
Figure~\ref{fig:2}(a) gives the total current $J_t$, coherent current $J_c$,
and incoherent current $J_{nc}$ versus $\Gamma_{d}$
for the two-terminal graphene nanoribbon system.
Here the methods for calculating $J_t$, $J_c$, and $J_{nc}$ are the same
as in reference~\cite{PhysRevB.77.115346}.
In the absence of dephasing ($\Gamma_{d}=0$), the incoherent current $J_{nc}$ is zero
and the coherent current $J_{c}$ is equal to the total current $J_{t}$.
With the increase of $\Gamma_{d}$, $J_{c}$ decreases and $J_{nc}$ increases,
while the total current $J_{t}$ is unchanged.
When the coherent component is just equal to the incoherent component,
the length $LT$ is just equal to the phase coherence length $L_\varphi$,
then the dependence between $L_\varphi$ and $\Gamma_{d}$ can be obtained.
Figure~\ref{fig:2}(b) shows $\Gamma_{d}$ versus $L_\varphi$ for different magnetic fields $B$.
When $\Gamma_{d}\rightarrow0$, $L_\varphi\rightarrow\infty$.
In this case, the device size $(N,LT)\ll L_\varphi$
thus electrons keep perfect quantum coherence.
With the increase of $\Gamma_{d}$, $L_\varphi$ decreases monotonically, thus electrons perfect quantum coherence is destroyed gradually.
For example, when $\Gamma_{d}=0.05$, $L_\varphi=63$ $(43)$ for $B=0.01$ $(0.02)$, which is similar to $N$. When $\Gamma_{d}=0.1$, $L_\varphi=32$ $(22)$
for $B=0.01$ $(0.02)$, which is smaller than $N$. And when $\Gamma_{d}=0.5$, $L_\varphi=7$ $(4)$ for $B=0.01$ $(0.02)$, which is much smaller than $N$.

\section{QHE appears if the lead-central region coupling is perfect }

\begin{figure}
      \includegraphics[width=0.8\textwidth]{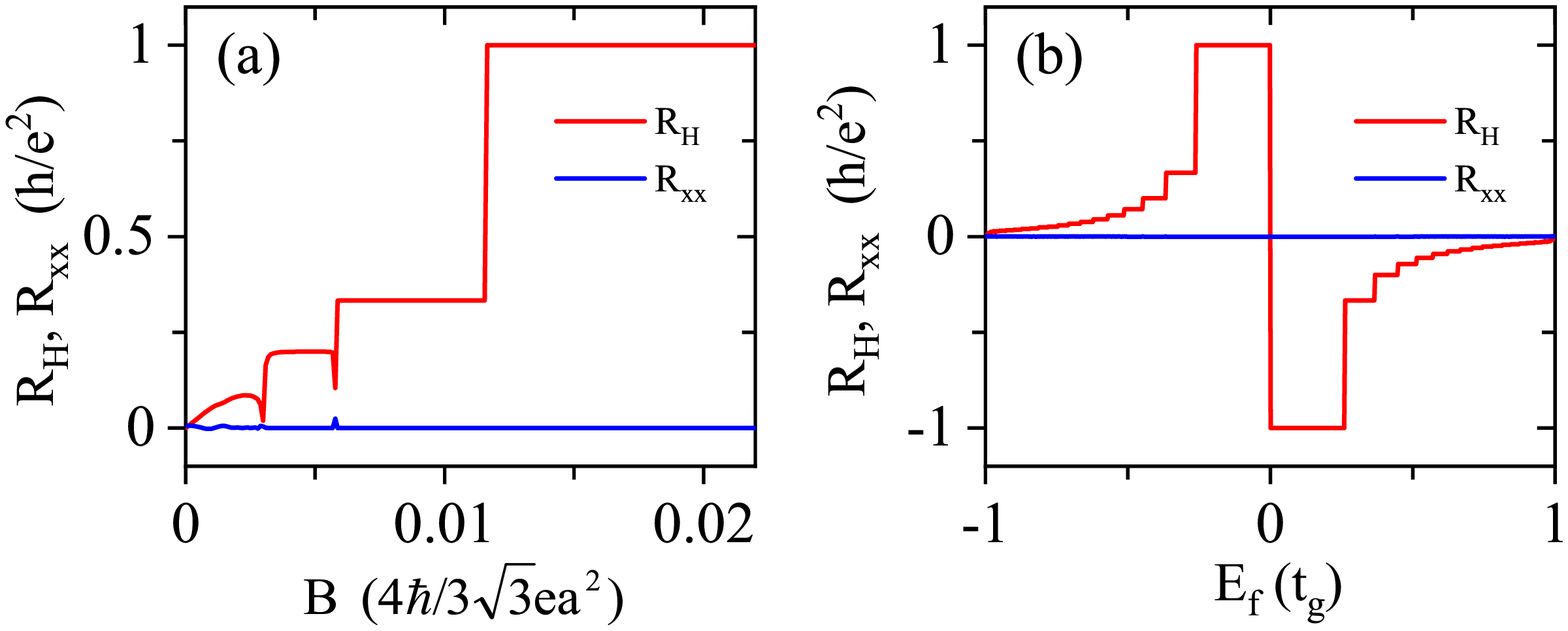}
      \centering
      \caption{Hall resistance $R_{H}$ and longitudinal resistance $R_{xx}$
    vs magnetic field $B$ in (a), Fermi energy $E_{f}$ in (b) with perfect lead-central region coupling.
    The coupling strength $t_{c1}=t_{g}$, $E_{f}=-0.2t_{g}$ in (a), $B=0.02$ in (b).}
      \label{fig:3}
    \end{figure}

In this section, we consider the situation that the left/right leads are perfectly coupled to central region
in perfectly quantum coherent systems ($\Gamma_d=0$).
First, this perfect coupling requires the left/right leads and central region are the same material (here graphene is considered, see figure~\ref{fig:1}(a)).
Next, $t_{c1}$, the coupling strength between left/right leads and central region,
is also required to be equal to $t_g$, the nearest neighbor carbon-carbon coupling strength in central region.
When the above two conditions are satisfied simultaneously, the lead-central region coupling is perfect and electrons can be fully transmitted without reflection
at the interfaces between left/right leads and central region.
Figures~\ref{fig:3}(a) and (b) show the Hall resistance $R_{H}$ and longitudinal resistance $R_{xx}$
versus magnetic field $B$ and Fermi energy $E_{f}$ with $t_{c1} = t_g$, respectively.
In figures~\ref{fig:3}(a) and (b), the dephasing strength $\Gamma_{d}=0$, thus the phase coherence length $L_{\varphi} \rightarrow \infty$ (see figure~\ref{fig:2}(b)).
In this case, $L_{\varphi} \gg (N, LT) \gg l_{B}$,
so electrons keep perfect quantum coherence,
and the upper and lower chiral edge states are well separated.
From figure~\ref{fig:3}(a), we can see the expected QH plateaus at $R_{H}=(1/\nu)h/e^2$ appear, with filling factors $\nu=1,3,5\dots$,
and the longitudinal resistance $R_{xx}=0$.
Similarly, we can also clearly see the QH plateaus and corresponding zero longitudinal resistance in figure~\ref{fig:3}(b).
So the QHE can appear in perfectly quantum coherent systems
if the lead-central region coupling is perfect,
which has been shown in many literature~\cite{PhysRevB.104.115411,PhysRevLett.108.166602,PhysRevB.87.235405,PhysRevB.81.245417}.

\section{QHE disappears if the lead-central region coupling is imperfect}

\begin{figure}
\includegraphics[width=0.8\textwidth]{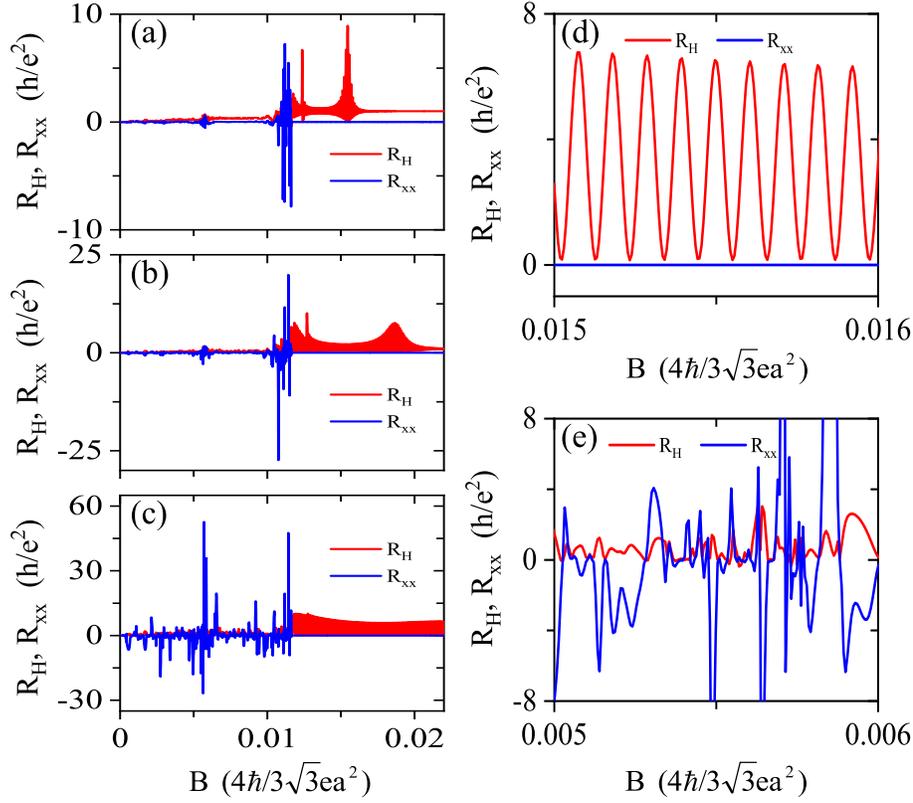}
\centering
\caption{(a-c) Hall resistance $R_{H}$ and longitudinal resistance $R_{xx}$ vs magnetic field $B$
for the case that the lead and central region are the same material but the coupling between them is imperfect.
The parameters are $\Gamma_d=0$, $E_{f}=-0.2t_{g}$, and $t_{c1}=0.5t_{g}$ in (a), $0.3t_{g}$ in (b), and $0.1t_{g}$ in (c). (d) and (e) are zoomed-in figures of (c).}
\label{fig:4}
\end{figure}

In this section, we study the situation that the left/right leads are imperfectly coupled to central region.
We still consider the perfectly quantum coherent system with
the dephasing strength $\Gamma_d=0$ ($L_{\varphi} \rightarrow \infty$). First, we study the case where the left/right leads and central region are the same material but the coupling between them is imperfect, i.e., $t_{c1}\neq t_g$.
Figures~\ref{fig:4}(a-c) show the Hall resistance $R_{H}$ and
longitudinal resistance $R_{xx}$ versus
the magnetic field $B$ with $t_{c1}\neq t_g$.
For comparison, the perfect coupling case with $t_{c1}= t_g$
has been shown in figure~\ref{fig:3}(a).
When the coupling between the left/right leads and central region is imperfect, reflection generally occurs at their interfaces.
Notice that electrons still propagate along the chiral edge states unidirectionally without backscattering in this case
(see the red solid lines in figure~\ref{fig:1}(a)),
so the QH plateau of $R_{H}$ and zero longitudinal resistance $R_{xx}$ should appear intuitively.
However, from figures~\ref{fig:4}(a-c), we can see
that both $R_H$ and $R_{xx}$
strongly oscillate with magnetic field $B$.
The QH plateau of $R_H$ and zero $R_{xx}$ disappear
except for the filling factor $\nu=1$.
When $\nu=1$, the longitudinal resistance $R_{xx}$ is zero,
while Hall resistance $R_H$ oscillates with $B$ periodically
with a period $1.1\times10^{-4}$
(see figure~\ref{fig:4}(d)), which is slightly larger than $\frac{2\pi\phi_0}{S}=8.8\times10^{-5}$,
with $S=\frac{3\sqrt{3}a^2}{2}(\frac{N}{2}-1)\times LT$ the area of the central region.
When $\nu>1$ (e.g., $\nu=3,5,\dots$), $R_{xx}$ and $R_{H}$
strongly oscillate with $B$,
and $R_{xx}$ even can be negative (see figure~\ref{fig:4}(e)).
Besides, when $t_{c1}$ deviates more from $t_g$,
the lead-central region coupling is much worse,
then $R_H$ and $R_{xx}$ oscillate more strongly.
$R_{H}$ and $R_{xx}$ versus Fermi energy $E_{f}$ has similar phenomena,
i.e., the QH plateau of $R_{H}$ disappears and $R_{xx}$ is nonzero for $\lvert\nu\rvert>1$.
These results indicate that QHE disappears
in perfectly quantum coherent systems with imperfect lead-central region coupling.

\begin{figure}
\includegraphics[width=0.8\textwidth]{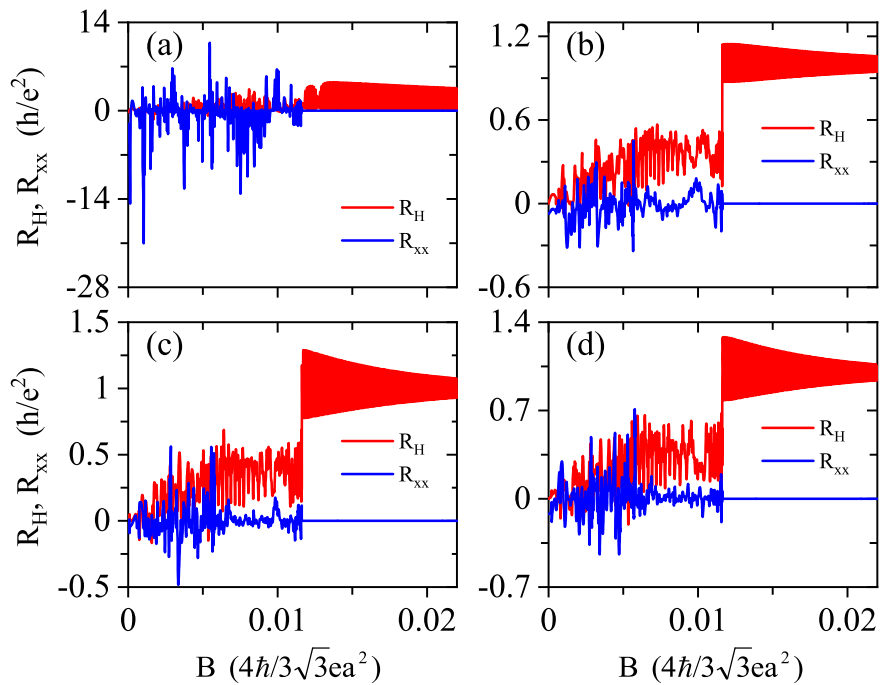}
\centering
\caption{Hall resistance $R_{H}$ and longitudinal resistance $R_{xx}$ vs magnetic field $B$ for the case that the leads and central region are different materials. The parameters are $\Gamma_d=0$, $E_{f}=-0.2t_{g}$,
and $t_{c2}=0.1t_{g}$ in (a), $0.5t_{g}$ in (b), $0.8t_{g}$ in (c), and $t_{g}$ in (d).}
\label{fig:5}
\end{figure}

In experiments, leads and central region
are usually not the same material~\cite{Nature.438.197,Nature.438.201},
so the coupling between them is imperfect
and electrons can be reflected at their interfaces. To better simulate the experiment, below we study the case
where the central region is graphene and
the left/right leads are normal metals (see figure~\ref{fig:1}(b)).
Figure~\ref{fig:5} shows the Hall resistance $R_{H}$ and
longitudinal resistance $R_{xx}$ versus the magnetic field $B$
for different lead-central region coupling strengths $t_{c2}$.
We can see that $R_{H}$
has no plateaus and $R_{xx}$ is nonzero
except for the filling factor $\nu=1$.
When $\nu=1$, $R_{H}$ oscillates with $B$ periodically
and the corresponding $R_{xx}$ is zero.
When $\nu>1$, both $R_{H}$ and $R_{xx}$ strongly oscillate with $B$.
Besides, $R_{H}$ and $R_{xx}$ versus $E_{f}$ have the similar behaviors, i.e., $R_{H}$ has no plateaus and $R_{xx}$ is nonzero except for $\lvert\nu\rvert=1$.
These results are the same as those for imperfect couplings in figure~\ref{fig:4}, because it's difficult to make the coupling between the central region and the leads
perfect and unreflected if they are different materials.
Therefore, the QHE always disappears in perfectly quantum coherent systems with imperfect lead-central region coupling.

The above results show that QHE disappears in perfectly quantum coherent systems with imperfect lead-central region coupling, which are counterintuitive.
We now discuss the cause of disappearance of the QHE
by using a simple ballistic transport picture (see Appendix A for the details). QHE requires lack of scattering
between the upper and lower voltage probes
(i.e., lead 2 (or 3) and lead 6 (or 5)). Under the perpendicular magnetic field, electrons propagate along the chiral edge states unidirectionally~\cite{Nature.461.772}. When device size is much larger than magnetic length
($N \gg l_B$), the upper and lower chiral edge states are
well separated in space without backscattering.
So electrons can't propagate directly from the upper edge to the lower edge. However, if the lead-central region coupling is imperfect,
electrons can be reflected at their interfaces. In this case, when electrons propagate along
the upper edge states and reach to the right
interface between lead 4 and central region,
some of them will be reflected to the lower edge states.
Similarly, when electrons continue to propagate along
the lower edge states and reach to the left interface
between lead 1 and central region, some of them will
be reflected to the upper edge states again.
Then, a closed propagating loop forms in which electrons can
keep turning in circles (see the red solid lines in figure~\ref{fig:1}).
As a result, electrons can propagate between the upper
and lower voltage probes through the closed loop,
which leads to the disappearance of QHE.
When the filling factor $\nu=1$, there is one chiral edge state in the closed loop,
leading to a periodic oscillation of Hall resistance $R_{H}$
(see Appendix A). Because of the extended edge state, the period is slightly larger than the theoretical one $\frac{2\pi\phi_0}{S}$. When $\nu>1$, the number of edge states is larger than 1, thus
the interference between these edge states leads to random oscillations
of $R_H$ and $R_{xx}$. Therefore, QHE disappears in perfectly quantum coherent systems, unless the lead-central region coupling is perfect
without reflection at their interfaces.

\section{Dephasing effect promotes the appearance of QHE in graphene system}

In the above section, we show that QHE disappears in perfectly quantum coherent systems,
unless the lead-central region coupling is perfect
without reflection. However, the QHE can be observed in experiments, although the lead-central region coupling is usually imperfect. To investigate which mechanism promotes the appearance of the QHE, we study the influence of dephasing effect on the QH plateaus in this section. In fact, dephasing inevitably exists
because of electron-phonon interactions, electron-electron interactions, etc.~\cite{PhysRevB.78.045322,PhysRevA.41.3436}, so the perfect quantum coherence is generally destroyed. Considering the leads and central region are usually different materials in experiment, here we study the case
where the central region is hexagonal graphene and
the left/right leads are normal square-lattice metals
(see figure~\ref{fig:1}(b)).

\begin{figure}
  \includegraphics[width=1\textwidth]{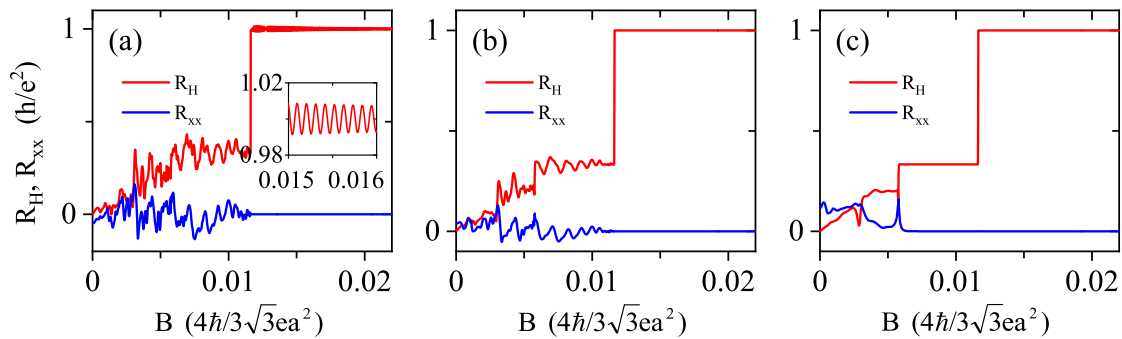}
  \centering
  \caption{Hall resistance $R_{H}$ and longitudinal resistance $R_{xx}$
  vs magnetic field $B$ for different dephasing strengths $\Gamma_{d}$.
  $\Gamma_{d}=0.05$ in (a), 0.1 in (b), and 0.5 in (c). The coupling strength $t_{c2}=0.1t_{g}$ and Fermi energy $E_{f}=-0.2t_{g}$.
  For comparison, the perfect coherence case with $\Gamma_d= 0$
  has been shown in figure~\ref{fig:5}(a).
  The inset is zoomed-in figure in the corresponding main figure (a).}
  \label{fig:6}
  \end{figure}

\begin{figure}
  \includegraphics[width=1\textwidth]{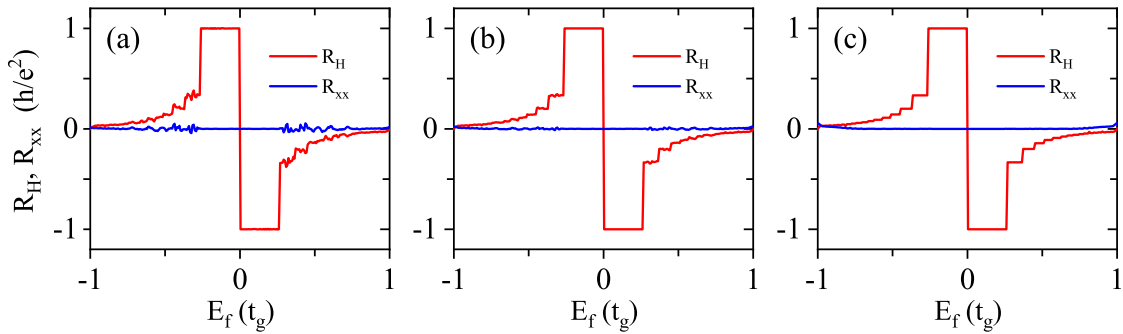}
  \centering
  \caption{Hall resistance $R_{H}$ and longitudinal resistance $R_{xx}$
  vs Fermi energy $E_f$ for different dephasing strengths $\Gamma_{d}$.
  $\Gamma_{d}=0.05$ in (a), 0.1 in (b), and 0.5 in (c), The coupling strength $t_{c2}=0.1t_{g}$ and magnetic field $B=0.02$.}
  \label{fig:7}
  \end{figure}

Figure~\ref{fig:6} gives the Hall resistance $R_{H}$ and
longitudinal resistance $R_{xx}$ versus magnetic field $B$
in the presence of dephasing ($\Gamma_{d}\neq 0$). With the increase of $\Gamma_d$, the quantum coherence becomes worse and the phase coherence length $L_{\varphi}$ decreases. From figure~\ref{fig:6}, we can see that the QH plateau appears gradually with increasing $\Gamma_{d}$. For a small dephasing strength $\Gamma_{d}=0.05$, $L_{\varphi}$ is similar to the device size $N$ ($L_{\varphi}\sim N$).
In this case, $R_{H}$ and $R_{xx}$ still oscillate with $B$, but the amplitude reduces obviously
compared to the case of $\Gamma_{d}=0$ (see figures~\ref{fig:6}(a) and \ref{fig:5}(a)). Since $L_{\varphi}\sim N$ at this time, this can also be regarded as a mesoscopic effect.
Continue to increase $\Gamma_{d}$, $L_{\varphi}$ is further reduced. When $L_\varphi$ is smaller than $N$ but larger than the magnetic length $l_B$ ($N>L_\varphi>l_B$), 
the QH plateaus appear and keep well, see figures~\ref{fig:6}(b) and (c). 
When $\Gamma_{d}=0.1$, the QH plateau with $R_{H}=1$ appears (figure~\ref{fig:6}(b)). 
When $\Gamma_{d}=0.5$, the QH plateau with $R_{H}=1, 1/3$ appear (figure~\ref{fig:6}(c)). Usually, the size
commonly used in experiments to realize the QHE is also in this range, $N>L_\varphi>l_B$~\cite{Nature.438.197,Nature.438.201}.
In figure~\ref{fig:7}, we study $R_{H}$
and $R_{xx}$ versus the Fermi energy $E_{f}$
in the presence of dephasing. Similarly, the QH plateaus as well as the zero longitudinal resistance appear gradually with increasing $\Gamma_{d}$.
In figures~\ref{fig:7}(a) and (b), the zero $R_{xx}$ and the QH resistance $R_{H}$ with filling factor $\nu=\pm1$ appear for $\Gamma_{d}=0.05$ and $0.1$. And when $\Gamma_{d}=0.5$, the zero $R_{xx}$ and the QH resistance $R_{H}$ with filling factor $\nu=\pm1, \pm3, \pm5,\cdots$ appear (figure~\ref{fig:7}(c)).
The above results indicate that dephasing effect
promotes the appearance of QHE.

Now, let's analyze the reason that why dephasing effect
promotes the appearance of QHE. When strong dephasing is added in the system,
$L_{\varphi}<N$ thus perfect quantum coherence is destroyed.
In dephasing regions, electrons propagating along the chiral edge states will lose phase memories and occur dissipation~\cite{PhysRevB.104.115411}.
In fact, dissipation in the chiral edge states has been observed
in a recent experiment~\cite{Nature.575.628}.
The distribution of electrons will finally
tend to equilibrium in dephasing regions~\cite{PhysRevB.104.115411}.
In this case, although the lead-central region coupling
is still imperfect and electrons can be reflected at their interfaces,
there is no coherent scattering between the upper and lower voltage probes,
then the QHE appears. In Appendix B, we analytically show
the appearance of QHE based on the above simple ballistic transport picture
(see Appendix B for more details).

\section{Dephasing effect promotes the appearance of QHE in disordered graphene system}

In the above section, we have shown dephasing effect promotes the appearance of QHE in the system without disorder for imperfect lead-central region coupling. In our numerical calculation, we mainly study the resistance $R$ (including $R_H$ and $R_{xx}$) versus magnetic field $B$ and Fermi energy $E_f$. On the curve of $R-B$, $E_f$ is fixed in our calculation, but what is actually fixed experimentally is the electron density $n_s$. Similarly, on the curve of $R-E_f$, the horizontal coordinate is $E_f$ in our calculation, while it is $n_s$ in experiments. At the thermodynamic limit, the ratio of the density of states of the edge and bulk tends to zero without disorder, thus the width of the Hall plateau will tend to zero both on the curve of $R_H-n_s$ and $R_H-B$. So disorder is indeed essential to realize the QHE in this case. Considering this situation, we next study the influence of dephasing effect on the QH plateaus in the presence of disorder. Here we consider the Anderson on-site disorder which only exists in the central region. In the presence of disorder, the on-site energy $\epsilon_{\bf i} =\epsilon_{0}+\omega_{\bf i}$ in Hamiltonian $H_{\textrm G}$, where the disordered energy $\omega_{\bf i}$ is uniformly distributed in the range $[-D/2, D/2]$ with the disorder strength $D$~\cite{PhysRevB.81.245417}.
Here we only consider a single disorder configuration instead of taking the average over a large number of disorder 
configurations. For the different disorder configurations, the results are qualitatively identical.

\begin{figure}
	\includegraphics[width=0.8\textwidth]{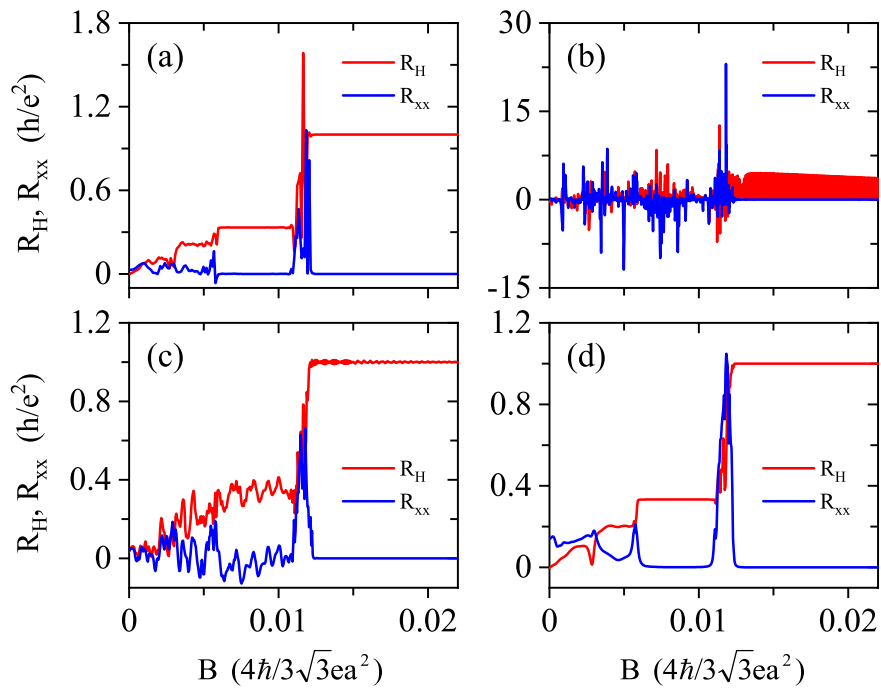}
	\centering
	\caption{Hall resistance $R_H$ and longitudinal resistance $R_{xx}$ vs magnetic field $B$ in the system with disorder, and the disorder strength $D=0.5$. The lead central region coupling is perfect in (a) with $t_{c1}=t_{g}$ and imperfect in (b)-(d) with $t_{c2}=0.1t_{g}$. And the dephasing strength $\Gamma_d=0$ in (a), (b), 0.05 in (c), and 0.5 in (d).}
	\label{fig:8}
\end{figure}

Figure~\ref{fig:8} shows the influence of dephasing effect on the QH plateaus in the presence of disorder. In figure~\ref{fig:8}(a), the lead-central region coupling is perfect and electrons keep perfect quantum coherence ($L_\varphi \gg N$), we can see that the QH plateau appears in this case. While when the lead-central region coupling is imperfect, the Hall resistance $R_H$ and the longitudinal resistance $R_{xx}$ oscillate strongly with magnetic field $B$ in perfectly quantum coherent system ($L_\varphi \gg N$), see figure~\ref{fig:8}(b). This indicates that the QHE disappears in perfectly quantum coherent systems if the lead-central region coupling is imperfect. Figures~\ref{fig:8}(c) and (d) show $R_H$ and $R_{xx}$ versus $B$ in a system with dephasing for imperfect lead-central coupling case. We can see that the QH plateau appears gradually after introducing dephasing. In figure~\ref{fig:8}(c), the dephasing strength $\Gamma_d$ is relatively small. $R_H$ and $R_{xx}$ still oscillate with $B$, but the amplitude reduces obviously, and the maximum value of $R_H$ is close to 1. For a larger $\Gamma_d$ in figure~\ref{fig:8}(d), the QH plateaus with values of 1 and 1/3 appear. From the above analysis, we can find that dephasing effect promotes the appearance of QH plateaus in the system with disorder, which is similar to the result obtained in the system without disorder.

In the results above, we have shown that dephasing effect promotes 
the appearance of QH plateaus regardless of presence or absence of disorder.
With the increase of the dephasing strength $\Gamma_d$ from $0$ to $\infty$,
the phase coherence length $L_\varphi$ gradually reduces from $\infty$ to $0$,
and QH plateaus gradually appear with $L_\varphi$ from $L_\varphi \gg N$ 
through $L_\varphi \sim N$ to $N\gg L_\varphi \gg l_B$.
The QH plateaus well form at the region of $N\gg L_\varphi \gg l_B$.
In general, the size commonly used in experiments to realize the QHE
is also in this range.
If the dephasing strength is further increased, $L_\varphi$ will become much smaller. 
When $L_\varphi$ is smaller than $l_B$ or the localization length $\xi$ 
($L_\varphi<l_B$ or $\xi$), the localization of electrons is affected 
thus the QH plateau shortens gradually~\cite{RevModPhys.67.357,PhysRevLett.61.1297}.

\begin{figure}
  \includegraphics[width=1\textwidth]{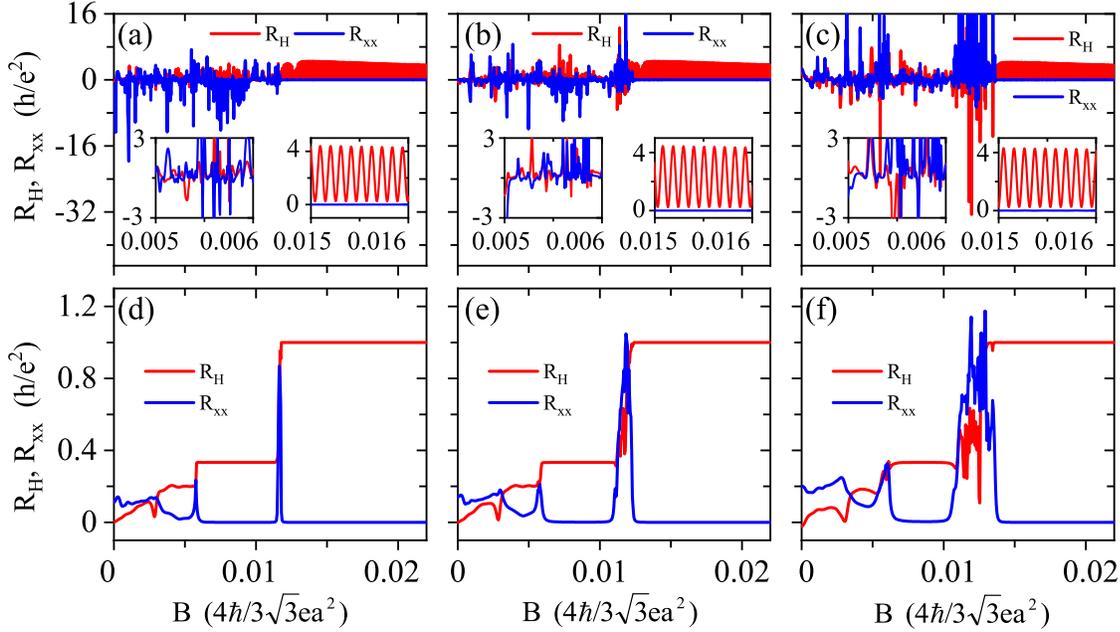}
  \centering
  \caption{Hall resistance $R_{H}$ and longitudinal resistance $R_{xx}$ vs magnetic field $B$ for different disorder strengths $D$. $\Gamma_{d}=0$ in (a)-(c) and 0.5 in (d)-(f). $D=0.1$ in (a), (d), 0.5 in (b), (e), and 1.0 in (c), (f). The insets are zoomed-in figures in the corresponding main figures (a), (b), and (c), respectively. Here figures~\ref{fig:9}(b) and (e) are the same as figures~\ref{fig:8}(b) and (d), respectively.}
  \label{fig:9}
\end{figure}

In fact, disorder inevitably exists since it can be induced by nonmagnetic impurities or random potential difference of the substrate in experiment~\cite{Nanolett.9.1973}. For example, disorder in graphene $p$-$n$ junction can result in several extra conductance plateaus~\cite{Science.317.638}. To prove that the appearance of QH plateaus here is indeed promoted by dephasing rather than disorder, we now turn our attention to the influence of disorder effect on the QHE. Figures~\ref{fig:9}(a)-(c) plot the Hall resistance $R_{H}$
and longitudinal resistance $R_{xx}$ versus
magnetic field $B$ for different disorder strengths at $\Gamma_{d}=0$.
For all disorder strengths $D$, $R_{H}$
has no plateaus and $R_{xx}$ is nonzero,
and both of them oscillate strongly with $B$, except for $\nu=1$.
For $\nu=1$, $R_{H}$ oscillates periodically and $R_{xx}$ is zero
(see the insets in figures~\ref{fig:9}(a)-(c)).
These results are the same as figure~\ref{fig:5} with $D=0$,
indicating that disorder can't promote the appearance of QHE.
On the other hand, in the presence of a large $\Gamma_{d}$,
the QH plateaus and zero $R_{xx}$ appear in the system without disorder ($D=0$),
see figure~\ref{fig:6}(c).
Figures~\ref{fig:9}(d)-(f) show $R_{H}$ and $R_{xx}$ versus
$B$ for different disorder strengths $D$
in the presence of a large $\Gamma_{d}$.
For small $D$, the QH plateaus and zero $R_{xx}$ hardly change (figure~\ref{fig:9}(d)). However, they are destroyed gradually by further increasing $D$. And those near the plateau transition region are destroyed first (figures~\ref{fig:9}(e) and (f)). In consequence, strong disorder is harmful to the QH plateaus~\cite{PhysRevLett.78.318,PhysRevB.73.233406,PhysRevB.59.8144,PhysRevLett.117.056802,PhysRevLett.76.975,RevModPhys.80.1355}, but dephasing is beneficial to them.

\section{Dephasing effect promotes the appearance of QHE in 2DEG system}

\begin{figure}
  \includegraphics[width=0.8\textwidth]{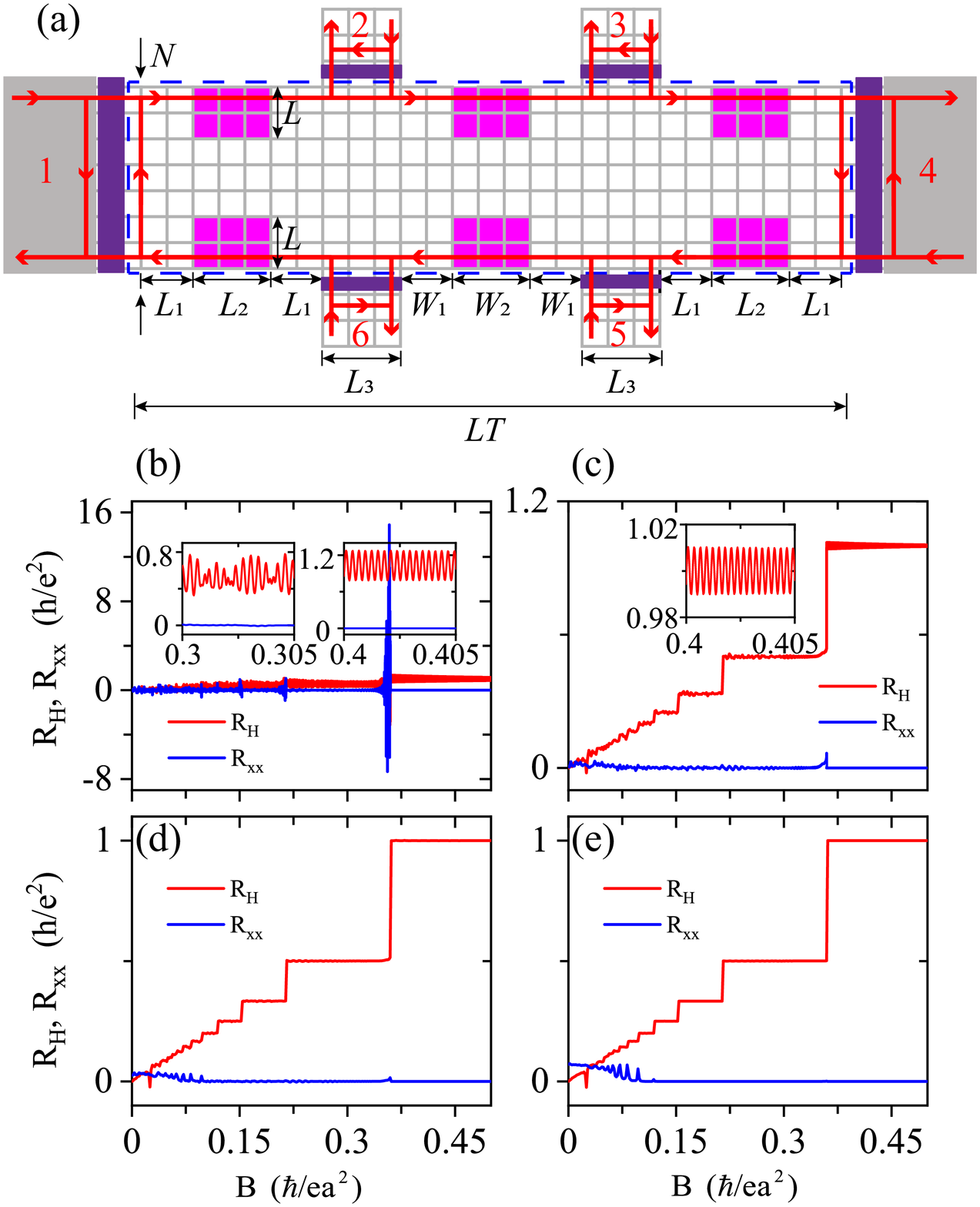}
  \centering
  \caption{(a) Illustration of a six-terminal Hall bar device consists of
	2DEG coupled to metal leads 1 and 4, with dephasing added in the magenta-site regions.
	The box with blue dotted lines is the central region. The red solid lines with arrows represent the propagation of electrons along the edge states. The purple areas represent the interfacial resistances
    when the lead-central region coupling is imperfect.
	In this diagram, the device sizes are $N=8$, $L=3$, $L_1=W_1=2$, $L_2=W_2=4$, and $L_3=2$.
	The illustration of the six-terminal QAHI device is the same as this figure.
  (b)-(e) show the Hall resistance $R_{H}$ and longitudinal resistance $R_{xx}$ vs magnetic field $B$ for different dephasing strengths
  $\Gamma_{d}$. $\Gamma_{d}=0$ in (b),
  0.01 in (c), 0.02 in (d), 0.05 in (e).
  The insets are zoomed-in figures in the corresponding main figures (b) and (c), respectively. In our calculations, the device sizes are $L_1=W_1=40$, $L_2=W_2=120$, $L_3=2$, and $L=10$.}
  \label{fig:10}
  \end{figure}

The QHE is first discovered in 2DEG and later realized in graphene~\cite{PhysRevLett.45.494,Nature.438.197,Nature.438.201}.
In the discussions above,
we have shown dephasing effect promotes the appearance of QHE in graphene system with imperfect lead-central region coupling.
In this section, we show that those results are the same in 2DEG system.
In the tight-binding representation, the Hamiltonian of a six-terminal Hall bar device (figure~\ref{fig:10}(a))
consisting of 2DEG coupled to the left and right metal leads can be expressed as
\begin{eqnarray}\label{eq:5}
H=H_{\rm 2DEG}+H_{\rm M3}+H_{\rm C3}+H_{\rm D3},
\end{eqnarray}
where $H_{\rm 2DEG}=\sum_{\bf i}(\epsilon_{\bf i}-E_{f})d_{\bf i}^{\dag}d_{\bf i}-\sum_{\langle{\bf ij}\rangle}
t_{s}e^{i\phi_{\bf ij}}d_{\bf i}^{\dag}d_{\bf j}$ is the Hamiltonian of 2DEG. $d_{\bf i}^\dag$ ($d_{\bf i}$) is the creation (annihilation) operator
at site ${\bf i}$. $\epsilon_{\bf i}=0$ is the on-site energy, $t_{s}=1$ meV is the hoping energy, and $E_{f}=-3t_{s}$ is the Fermi energy. Note that the 2DEG model is a square-lattice (figure~\ref{fig:10}(a)), while the graphene model is a hexagonal-lattice (figure~\ref{fig:1}(b)). For the left/right leads, we also use a square-lattice Hamiltonian
$H_{\textrm M3}=\sum_{\bf i}\epsilon_{3}b_{\bf i}^{\dag}b_{\bf i}-\sum_{\langle{\bf ij}\rangle}t_{m3}b_{\bf i}^{\dag}b_{\bf j}$
with $\epsilon_{3}=3t_{s}$ the on-site energy, $t_{m3}=t_{s}$ the hoping energy. The coupling between the metal leads and 2DEG is described by the Hamiltonian $H_{\rm C3}=-t_{c3}\sum_{\langle{\bf ij}\rangle}b_{\bf i}^{\dag}d_{\bf j}+\textrm{H.c.}$, with $t_{c3}=0.3t_{s}$ the coupling strength. Under this coupling strength, the lead-central region coupling is imperfect. The Hamiltonian of the virtual leads and their couplings to the corresponding sites is described as $H_{\rm D3}=\sum_{{\bf i},k}\epsilon_{k}a_{{\bf i}k}^{\dag}a_{{\bf i}k}
 +(t_{k}a_{{\bf i}k}^{\dag}d_{\bf i}+\textrm{H.c.})$.

Figure~\ref{fig:10}(b) gives the Hall resistance $R_{H}$ and the longitudinal resistance $R_{xx}$
as functions of magnetic field $B$ in 2DEG system with $\Gamma_{d}=0$. When $\Gamma_{d}=0$,
$L_{\varphi}\rightarrow\infty$. In this case, $L_{\varphi}\gg(N,LT)\gg l_{B}$, electrons keep perfect quantum coherence and the chiral edge states are well separated in space, thus the QH plateaus should appear intuitively.
However, no QH plateaus appear in this case. $R_{H}$ and $R_{xx}$ strongly oscillates with $B$ except for the filling factor $\nu=1$ (see the left inset in figure~\ref{fig:10}(b)). When $\nu=1$,
$R_{H}$ oscillates with $B$ periodically with a period $3\times10^{-4}$
(see the right inset in figure~\ref{fig:10}(b)),
which is slightly larger than $\frac{2\pi\phi_0}{S} = 2.67\times10^{-4}$ (with a unit of $\frac{\hbar}{ea^2}$),
where $S=a^2(N-1)(LT-1)$ is the area of the central region. Figures~\ref{fig:10}(c)-(e) show $R_{H}$ and $R_{xx}$ versus $B$ in the presence of dephasing ($\Gamma_{d} \neq 0)$.
When $\Gamma_{d}$ is small, $R_{H}$ still oscillates with $B$, but the amplitude decreases obviously (see figure~\ref{fig:10}(c)). With the further increase of $\Gamma_{d}$, the QHE with QH plateau at $R_{H}=(1/\nu)h/e^2$ ($\nu=1,2,3,\cdots$) appears gradually, and the corresponding $R_{xx}$ is zero (figures~\ref{fig:10}(d) and (e)),
which means that dephasing effect promotes the appearance of QHE in 2DEG system with imperfect lead-central region coupling.

\section{Dephasing effect promotes the appearance of QAHE}

In addition to the QHE, the QH plateau
can also be observed in some two-dimensional magnetic topological insulators
without the magnetic field. This is called the famous QAHE and has been extensively investigated recently~\cite{Science.329.61,Science.340.167,NanoLett.20.7606,Science.367.895,Science.367.900}. Above we have shown: (i) QHE disappears in perfectly quantum coherent systems with imperfect lead-central region coupling. (ii) The dephasing effect promotes the appearance of QHE. Then one may wonder what will happen for the QAHE under the same condition. In the following, we study the QAHE in a six-terminal Hall bar device based on
a quantum anomalous Hall insulator (QAHI) (figure~\ref{fig:10}(a)).
The Hamiltonian can be expressed as
\begin{equation}\label{eq:7}
H=H_{\rm QAHI}+H_{\textrm M4}+H_{\textrm C4}+H_{\textrm D4},
\end{equation}
where $H_{\rm QAHI}$, $H_{\textrm M4}$, $H_{\textrm C4}$, and $H_{\textrm D4}$ are, respectively, the Hamiltonian of QAHI, metal leads,
coupling of QAHI and metal leads, B\"{u}ttiker's virtual leads and their couplings to central sites.
To describe the QAHI system, we consider a two-band effective
Hamiltonian $H_{\rm QAHI}({\bf p})=(m+Bp^2)\sigma_{z}+A(p_{x}\sigma_{x}+p_{y}\sigma_{y})-E_{f}\sigma_{0}$ with the momentum ${\bf p}$, Pauli matrices $\sigma_{x, y, z}$, identity matrix $\sigma_{0}$, Fermi energy $E_{f}$, and material parameters $A$, $B$, and $m$~\cite{PhysRevB.82.184516}.
The topological property of $H_{\rm QAHI}$ is determined by the sign of $m/B$.
For $m/B<0$, the QAHI state is topologically nontrivial with the Chern number $\mathcal{C}=1$, carrying one chiral edge mode.
Here we use the dimensionless parameters with $A=B=1$, and $m=-0.5$~\cite{PhysRevB.97.115452}. For the numerical calculations, $H_{\rm QAHI}$
can be further mapped into a square-lattice model in the tight-binding representation
\begin{eqnarray}\label{eq:8}
\mathcal{H}_{\rm QAHI}=\sum_{\bf i}[\psi_{\bf i}^{\dag}T_{0}\psi_{\bf i}+(\psi_{\bf i}^{\dag}T_{x}\psi_{\bf i+\delta{\bf x}}
+\psi_{\bf i}^{\dag}T_{y}\psi_{\bf i+\delta{\bf y}})+\textrm{H.c.}],
\end{eqnarray}
where $T_{0}=(m+4B\hbar^{2}/a^{2})\sigma_{z}-E_f\sigma_{0}$, $T_{x}=-(B\hbar^{2}/a^{2})\sigma_{z}-(iA\hbar/2a)\sigma_{x}$, and $T_{y}=-(B\hbar^{2}/a^{2})\sigma_{z}-(iA\hbar/2a)\sigma_{y}$. Here $\psi_{\bf i}=(c_{\bf i\uparrow}, c_{\bf i\downarrow})^{T}$, $c_{\bf i\sigma}$ and $c_{\bf i\sigma}^{\dag}$ are the annihilation and creation operators at site ${\bf i}$ with spin $\sigma=\uparrow$, $\downarrow$, respectively. $a$ is the lattice length and $\delta{\bf x}$ $(\delta{\bf y})$ is the unit cell vector along $x$ $(y)$ direction.
For the left and right metal leads, the Hamiltonian $H_{\rm M4}$ is
$H_{\textrm M4}=\sum_{\bf i}\epsilon_{4}b_{\bf i}^{\dag}b_{\bf i}-\sum_{\langle{\bf ij}\rangle}t_{m4}b_{\bf i}^{\dag}b_{\bf j}$,
where $\epsilon_{4}=3$ the on-site energy, $t_{m4}=1$ the hoping energy
and $b_{\bf i}= (b_{{\bf i}\uparrow},b_{{\bf i}\downarrow})^{T}$
the annihilation operator of the electrons. The coupling between the metal leads and the QAHI is described by the Hamiltonian
$H_{\textrm C4}=-t_{c4}\sum_{\langle{\bf ij}\rangle}b_{\bf i}^
{\dag}\psi_{\bf j}+\textrm{H.c.}$ with $t_{c4}=0.1t_{m4}$ the coupling strength. Under this coupling strength, the lead-central region coupling is imperfect. And $H_{\textrm D4}=\sum_{{\bf i},k}\epsilon_{k}a_{{\bf
i}k}^{\dag}a_{{\bf i}k}
+(t_{k}a_{{\bf i}k}^{\dag}\psi_{\bf i}+\textrm{H.c.})$ is the Hamiltonian of the virtual leads and their couplings to the central sites.

\begin{figure}
  \includegraphics[width=0.8\textwidth]{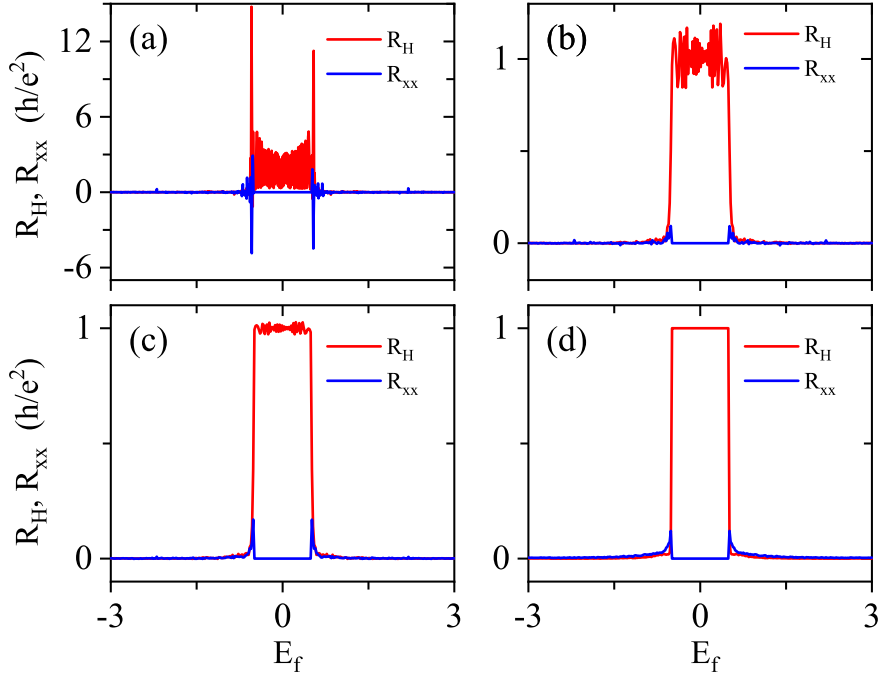}
  \centering
  \caption{Hall resistance $R_{H}$ and longitudinal resistance $R_{xx}$ vs Fermi energy $E_{f}$ for different $\Gamma_{d}$.
  $\Gamma_{d}=0$ in (a), 0.005 in (b), 0.01 in (c), and 0.05 in (d).}
  \label{fig:11}
  \end{figure}

Figure~\ref{fig:11} shows the Hall resistance $R_{H}$ and
longitudinal resistance $R_{xx}$ versus the Fermi energy $E_{f}$
for different dephasing strength $\Gamma_d$.
When $E_{f}$ is in the bulk bands ($|E_f|>|m|$),
$R_{H}$ and $R_{xx}$ are very small regardless of $\Gamma_d$.
Then we focus on the case that $E_f$ is in the bulk gap ($|E_f|<|m|$). In this case, $R_{xx}$ is always zero, so we mainly analyze the influence of dephasing effect on $R_{H}$. When $\Gamma_{d}=0$, electrons keep perfect quantum coherence, but no QH plateau appears and $R_{H}$ strongly oscillates with $E_{f}$ (see figure~\ref{fig:11}(a)). In the presence of a small $\Gamma_{d}$, $R_{H}$ still oscillates with $E_{f}$, but the amplitude decreases obviously (figures~\ref{fig:11}(b) and (c)). By further increasing $\Gamma_{d}$, the quantum coherence becomes worse while the QH plateau with value of
$1$ appears (figure~\ref{fig:11}(d)).
These results indicate that dephasing effect promotes the appearance of QAHE for imperfect lead-central region coupling.

\section{Conclusion}
We have investigated the influence of dephasing effect on the QH plateaus by using Green's function method and Landauer-B\"{u}ttiker formula.
We find that the QH plateaus disappear
in perfectly quantum coherent systems with imperfect lead-central region coupling.
However, when quantum coherence becomes worse in the presence of dephasing,
the QH plateaus and zero longitudinal resistance appear gradually,
which indicates that dephasing effect promotes the appearance of QHE.
Furthermore, these results are independent of specific materials
and similar results are also obtained for the QAHE.
Our results indicate that dephasing effect promotes the appearance of
QH plateaus, which provide new insight into
the influence of dephasing effect on topological systems.

\ack{This work was financially supported by National Key R and D Program of China (Grant No. 2017YFA0303301),
NSF-China (Grants No. 11921005 and No. 11874428),
and the Strategic Priority Research Program of Chinese
Academy of Sciences (Grant No. XDB28000000).}

\appendix
\section*{Appendix A: A ballistic transport picture to explain the disappearance of QH plateaus in perfectly quantum coherent systems}

\def\theequation{A\arabic{equation}}
\setcounter{equation}{0}

Based on the ballistic transport picture
(see the red solid lines in figure~\ref{fig:1}),
we explain the cause of disappearance of
the QHE in perfectly quantum coherent systems.
Without loss of generality, we assume the number of channels $M=1$
(i.e., the filling factor $\nu=1$). The scattering matrix at the interface
between lead 1 (or 4) and central region is $s1=\begin{bmatrix} -r' & t' \\ t' & r' \end{bmatrix}$, and scattering matrix at the interface between lead 2 (or 3, 5, 6) and central region is $s2=\begin{bmatrix} -r & t \\ t & r \end{bmatrix}$,
where $\lvert r \rvert^2+\lvert t \rvert^2=1$ and $\lvert r' \rvert^2+\lvert t' \rvert^2=1$~\cite{Datta}. When electrons start from lead 1 and propagate forward
along the chiral edge states, although there is no backscattering between the upper and lower edge states, electrons can be reflected at interfaces because of imperfect lead-central region coupling (see the purple areas in figure~\ref{fig:1}).
As a result, electrons can keep turning in circles and they will enter the other leads again and again through the closed loop.
Therefore, the scattering between the upper and lower voltage probes happens.
In this case, for electrons starting from lead 1 and entering lead 2,
the lowest-order (i.e., the zeroth-order) process is that electrons start from lead 1 and enter lead 2 directly, so the zeroth-order term is $t't$.
Then, when electrons start from lead 1 and make one turn around the closed loop into lead 2, they will accumulate a phase $\Phi=\Phi_{d}+\Phi_{B}$, where $\Phi_{d}$ is the dynamic phase and $\Phi_{B}=\frac{BS}{\phi_{0}}$ is the phase given by magnetic flux under the magnetic field, with $S$ the area of the central region, so the first-order term is $t'(rrr'rrr'e^{i\Phi})t$.
Similarly, when electrons make two turns around the closed loop into lead 2,
the second-order term is $t'(rrr'rrr'e^{i\Phi}rrr'rrr'e^{i\Phi})t$ and so on.
Therefore, the transmission coefficient $T_{2,1}$ can be expressed as
\begin{eqnarray}\label{eq:A1}
	T_{2,1}&=&\left\lvert t't+t'(rrr'rrr'e^{i\Phi})t+t'(rrr'rrr'e^{i\Phi}rrr'rrr'e^{i\Phi})t+\cdots \right\rvert^2 \nonumber\\
	&=&\lvert t't+t'tr^4r'^2e^{i\Phi}+t't(r^4r'^2e^{i\Phi})^2+\cdots \rvert^2 \nonumber\\
	&=&\lvert t't\{1+r^4r'^2e^{i\Phi}+(r^4r'^2e^{i\Phi})^2+\cdots\} \rvert^2 \nonumber\\
	&=&\frac{T'T}{\lvert 1-R^2R'e^{i\Phi} \rvert^2},
\end{eqnarray}
where $T=\lvert t \rvert^2$, $R=\lvert r \rvert^2$, $T'=\lvert t' \rvert^2$, $R'=\lvert r' \rvert^2$.

Similarly, the other transmission coefficients $T_{p,q}$ can be expressed as
\begin{align}\label{eq:A2}
T_{3,1}&=\frac{T'RT}{\lvert 1-R^2R'e^{i\Phi} \rvert^2}; & T_{4,1}&=\frac{T'^2R^2}{\lvert 1-R^2R'e^{i\Phi} \rvert^2}; & T_{5,1}&=\frac{T'R^2R'T}{\lvert 1-R^2R'e^{i\Phi} \rvert^2}; \nonumber  \\
T_{6,1}&=\frac{T'R^3R'T}{\lvert 1-R^2R'e^{i\Phi} \rvert^2}; & T_{1,1}&=\lvert -r'+\frac{T'R^2r'e^{i\Phi}}{1-R^2R'e^{i\Phi}} \rvert^2. \nonumber  \\
T_{3,2}&=\frac{T^2}{\lvert 1-R^2R'e^{i\Phi} \rvert^2}; &
T_{4,2}&=\frac{TRT'}{\lvert 1-R^2R'e^{i\Phi} \rvert^2}; &
T_{5,2}&=\frac{T^2RR'}{\lvert 1-R^2R'e^{i\Phi} \rvert^2}; \nonumber  \\
T_{6,2}&=\frac{T^2R^2R'}{\lvert 1-R^2R'e^{i\Phi} \rvert^2}; &
T_{1,2}&=\frac{TR^3R'T'}{\lvert 1-R^2R'e^{i\Phi} \rvert^2}; &
T_{2,2}&=\lvert -r+\frac{TRR're^{i\Phi}}{1-R^2R'e^{i\Phi}} \rvert^2. \nonumber  \\
T_{4,3}&=\frac{T'T}{\lvert 1-R^2R'e^{i\Phi} \rvert^2}; &
T_{5,3}&=\frac{T^2R'
}{\lvert 1-R^2R'e^{i\Phi} \rvert^2}; &
T_{6,3}&=\frac{T^2R'R
}{\lvert 1-R^2R'e^{i\Phi} \rvert^2}; \nonumber  \\
T_{1,3}&=\frac{TR'R^2T'
}{\lvert 1-R^2R'e^{i\Phi} \rvert^2}; &
T_{2,3}&=\frac{T^2R'^2R^2
}{\lvert 1-R^2R'e^{i\Phi} \rvert^2}; &
T_{3,3}&=\lvert -r+\frac{TRR're^{i\Phi}}{1-R^2R'e^{i\Phi}} \rvert^2. \nonumber  \\
T_{5,4}&=\frac{T'T}{\lvert 1-R^2R'e^{i\Phi} \rvert^2}; &T_{6,4}&=\frac{T'RT}{\lvert 1-R^2R'e^{i\Phi} \rvert^2}; &T_{1,4}&=\frac{T'^2R^2}{\lvert 1-R^2R'e^{i\Phi} \rvert^2}; \nonumber  \\
T_{2,4}&=\frac{T'R^2R'T}{\lvert 1-R^2R'e^{i\Phi} \rvert^2}; &
T_{3,4}&=\frac{T'R^3R'T}{\lvert 1-R^2R'e^{i\Phi} \rvert^2}; &
T_{4,4}&=\lvert -r'+\frac{T'R^2r'e^{i\Phi}}{1-R^2R'e^{i\Phi}} \rvert^2. \nonumber  \\
T_{6,5}&=\frac{T^2}{\lvert 1-R^2R'e^{i\Phi} \rvert^2}; &
T_{1,5}&=\frac{TRT'}{\lvert 1-R^2R'e^{i\Phi} \rvert^2}; &
T_{2,5}&=\frac{T^2RR'}{\lvert 1-R^2R'e^{i\Phi} \rvert^2}; \nonumber  \\
T_{3,5}&=\frac{T^2R^2R'}{\lvert 1-R^2R'e^{i\Phi} \rvert^2}; &
T_{4,5}&=\frac{TR^3R'T'}{\lvert 1-R^2R'e^{i\Phi} \rvert^2}; &
T_{5,5}&=\lvert -r+\frac{TRR're^{i\Phi}}{1-R^2R'e^{i\Phi}} \rvert^2. \nonumber  \\
T_{1,6}&=\frac{TT'}{\lvert 1-R^2R'e^{i\Phi} \rvert^2}; & T_{2,6}&=\frac{T^2R'}{\lvert 1-R^2R'e^{i\Phi} \rvert^2}; & T_{3,6}&=\frac{T^2R'R}{\lvert 1-R^2R'e^{i\Phi} \rvert^2}; \nonumber  \\
T_{4,6}&=\frac{TR'R^2T'}{\lvert 1-R^2R'e^{i\Phi} \rvert^2}; &
T_{5,6}&=\frac{T^2R'^2R^2}{\lvert 1-R^2R'e^{i\Phi} \rvert^2}; &
T_{6,6}&=\lvert -r+\frac{TRR're^{i\Phi}}{1-R^2R'e^{i\Phi}} \rvert^2.
\end{align}

The above transmission coefficients satisfy $\sum_{q}T_{p,q}=\sum_{q}T_{q,p}=M=1$.
When a small bias is applied between leads 1 and 4 with $V_1=-V_4=V$,
the current flows along the longitudinal direction.
The leads 2, 3, 5, 6 are set as voltage probes and their currents are zero.
Then, combing equation~(\ref{eq:4}) with all these transmission coefficients $T_{p,q}$ and boundary
conditions, we can get the transverse bias $V_2=V_3=\frac{T'}{1+R'}V$, $V_5=V_6=-\frac{T'}{1+R'}V$, and the longitudinal current $J_1=-J_4=\frac{e^2}{h}\frac{2T'(1-R^4R'^2)}{(1+R')\lvert 1-R^2R'e^{i\Phi} \rvert^2}V$.
Then the Hall resistance $R_H$ and the longitudinal resistance $R_{xx}$ can be expressed as
\begin{eqnarray}\label{eq:A3}
R_{H}&=&\frac{V_2-V_6}{J_1}=\frac{h}{e^2}\frac{\lvert 1-R^2R'e^{i\Phi} \rvert^2}{1-R^4R'^2}, \notag \\
R_{xx}&=&\frac{V_2-V_3}{J_1}=0.	
\end{eqnarray}

From equation~(\ref{eq:A3}), we can reproduce all of the above numerical results.
That is, when the lead-central region coupling is perfect ($R'=0$),
$R_{xx}$ is zero and $R_H$ has the quantized plateau.
On the other hand, when the lead-central region coupling
is imperfect ($R' \ne 0$),
the QH plateau disappears and the Hall resistance $R_H$
oscillates with $B$ periodically with a period $\frac{2\pi\phi_{0}}{S}$.

\section*{Appendix B: A ballistic transport picture to explain the appearance of QH plateaus in the presence of dephasing}

 \def\theequation{B\arabic{equation}}
\setcounter{equation}{0}

In this section, let's explain the cause of appearance of
the QHE in the presence of dephasing by using a simple ballistic transport picture. When strong dephasing is added in the system, perfect quantum coherence is destroyed.
In the dephasing regions {\uppercase\expandafter{\romannumeral1}}-{\uppercase\expandafter{\romannumeral6}} (see figure~\ref{fig:1}(b)),
although electrons propagate along
the chiral edge states unidirectionally without backscattering,
they will lose phase memories and occur dissipation,
then the distribution of electrons will tend to equilibrium~\cite{PhysRevB.104.115411}.
So we can assume that electrons in
dephasing regions {\uppercase\expandafter{\romannumeral1}}-{\uppercase\expandafter{\romannumeral6}} are in equilibrium
with the chemical potentials $eV_{{\rm \uppercase\expandafter{\romannumeral1}}-{\rm \uppercase\expandafter{\romannumeral6}}}$.
For simplicity, below we consider the channel number $M=1$
(i.e., the filling factor $\nu=1$).
When electrons start from lead 1 and reach to the left interface,
some of them will be reflected to lead 1,
which is represented by the matrix element $-r'$,
and the rest $t'$ will be transmitted
to dephasing region {\uppercase\expandafter{\romannumeral1}}.
Then, when electrons start from region {\uppercase\expandafter{\romannumeral1}}, there is a component of $t$ will enter lead 2, and the rest $r$ will be fully transmitted to dephasing region {\uppercase\expandafter{\romannumeral2}}
and so on.
Therefore, the transmission coefficients can be expressed as
\begin{align}\label{eq:B1}
T_{1,1}&=\lvert -r' \rvert^2=R'; & T_{\textrm {\uppercase\expandafter{\romannumeral1}},1}&=\lvert t' \rvert^2=T'; & T_{2,\textrm {\uppercase\expandafter{\romannumeral1}}}&=\lvert t \rvert^2=T; & T_{\textrm {\uppercase\expandafter{\romannumeral2}},\textrm {\uppercase\expandafter{\romannumeral1}}}&=\lvert r \rvert^2=R. \notag \\
T_{2,2}&=\lvert -r \rvert^2=R; & T_{\textrm {\uppercase\expandafter{\romannumeral2}},2}&=\lvert t \rvert^2=T; & T_{3,\textrm {\uppercase\expandafter{\romannumeral2}}}&=\lvert t \rvert^2=T; & T_{\textrm {\uppercase\expandafter{\romannumeral3}},\textrm {\uppercase\expandafter{\romannumeral2}}}&=\lvert r \rvert^2=R. \notag \\
T_{3,3}&=\lvert -r \rvert^2=R; & T_{\textrm {\uppercase\expandafter{\romannumeral3}},3}&=\lvert t \rvert^2=T; & T_{4,\textrm {\uppercase\expandafter{\romannumeral3}}}&=\lvert t' \rvert^2=T'; & T_{\textrm {\uppercase\expandafter{\romannumeral4}},\textrm {\uppercase\expandafter{\romannumeral3}}}&=\lvert r' \rvert^2=R'. \notag \\
T_{4,4}&=\lvert -r' \rvert^2=R'; & T_{\textrm {\uppercase\expandafter{\romannumeral4}},4}&=\lvert t' \rvert^2=T'; & T_{5,\textrm {\uppercase\expandafter{\romannumeral4}}}&=\lvert t \rvert^2=T; & T_{\textrm {\uppercase\expandafter{\romannumeral5}},\textrm {\uppercase\expandafter{\romannumeral4}}}&=\lvert r \rvert^2=R. \notag \\
T_{5,5}&=\lvert -r \rvert^2=R; & T_{\textrm {\uppercase\expandafter{\romannumeral5}},5}&=\lvert t \rvert^2=T; & T_{6,\textrm {\uppercase\expandafter{\romannumeral5}}}&=\lvert t \rvert^2=T; & T_{\textrm {\uppercase\expandafter{\romannumeral6}},\textrm {\uppercase\expandafter{\romannumeral5}}}&=\lvert r \rvert^2=R. \notag \\
T_{6,6}&=\lvert -r \rvert^2=R; & T_{\textrm {\uppercase\expandafter{\romannumeral6}},6}&=\lvert t \rvert^2=T; & T_{1,\textrm {\uppercase\expandafter{\romannumeral6}}}&=\lvert t' \rvert^2=T'; & T_{\textrm {\uppercase\expandafter{\romannumeral1}},\textrm {\uppercase\expandafter{\romannumeral6}}}&=\lvert r' \rvert^2=R'.
\end{align}

The unmentioned transmission coefficients $T_{p,q}=0$.
These transmission coefficients also satisfy $\sum_{q}T_{p,q}=\sum_{q}T_{q,p}=M=1$.
When a small bias is applied between leads 1 and 4 with $V_1=-V_4=V$, the current flows along the longitudinal direction. The leads 2, 3, 5, 6 are set as voltage probes and their currents are zero.
Besides, in regions {\uppercase\expandafter{\romannumeral1}}-{\uppercase\expandafter{\romannumeral6}},
the distribution of electrons is in equilibrium with
the chemical potentials being
$eV_{\rm{\uppercase\expandafter{\romannumeral1}}-\rm {\uppercase\expandafter{\romannumeral6}}}$,
and the current flowing in is equal to the current flowing out.
Then, by combing equation~(\ref{eq:4}) with all these above transmission coefficients $T_{p,q}$ and
boundary conditions, we can get the transverse bias $V_{\textrm {\uppercase\expandafter{\romannumeral1}}}=V_2=V_{\textrm {\uppercase\expandafter{\romannumeral2}}}=V_3=V_{\textrm {\uppercase\expandafter{\romannumeral3}}}=\frac{T'}{1+R'}V$, $V_{\textrm {\uppercase\expandafter{\romannumeral4}}}=V_5=V_{\textrm {\uppercase\expandafter{\romannumeral5}}}=V_6=V_{\textrm {\uppercase\expandafter{\romannumeral6}}}=-\frac{T'}{1+R'}V$, and the longitudinal current $J_1=-J_4=\frac{e^2}{h}\frac{2T'}{1+R'}V$. Then the Hall resistance $R_H$ and the longitudinal resistance $R_{xx}$ can be obtained straightforwardly
\begin{eqnarray}\label{eq:B2}
R_{H}&=&\frac{V_2-V_6}{J_1}=\frac{h}{e^2}, \notag \\
R_{xx}&=&\frac{V_2-V_3}{J_1}=0.	
\end{eqnarray}

From equation~(\ref{eq:B2}), we can see that
the longitudinal resistance $R_{xx}$ is zero
and the Hall resistance $R_H$ has the quantized value 1 with a unit of $h/e^2$,
although electrons can be reflected at the lead-central region interfaces
and propagate between the upper and lower voltage
probes through the closed loop.
Therefore, dephasing effect can promote the appearance of the QH plateaus.


\section*{References}

\begin{thebibliography}{99}

\bibitem{RevModPhys.81.109}
Castro Neto A H, Guinea F, Peres N M R, Novoselov K. S and Geim A K 2009 \textit{Rev. Mod. Phys.} {\bf 81} 109

\bibitem{RevModPhys.82.3045}
Hasan M Z and Kane C L 2010 \textit{Rev. Mod. Phys.} {\bf 82} 3045

\bibitem{PhysRevLett.45.494}
von Klitzing K, Dorda G and Pepper M 1980 \textit{Phys. Rev. Lett.} {\bf 45} 494

\bibitem{PhysRevLett.48.1559}
Tsui D C, St\"{o}rmer H L and Gossard A C 1982 \textit{Phys. Rev. Lett.} {\bf 48} 1559

\bibitem{PhysRevLett.50.1395}
Laughlin R B 1983 \textit{Phys. Rev. Lett.} {\bf 50} 1395

\bibitem{RevModPhys.58.519}
von Klitzing K 1986 \textit{Rev. Mod. Phys.} {\bf 58} 519

\bibitem{Prange}
Prange R E and Girvin S M 1990 \textit{The Quantum Hall Effect} (Berlin: Springer-Verlag) p 9

\bibitem{Nature.569.537}
Tang F \textit{et al} 2019 \textit{Nature} {\bf 569} 537

\bibitem{Nature.579.56}
Chen G \textit{et al} 2020 \textit{Nature} {\bf 579} 56

\bibitem{PhysRevB.23.5632}
Laughlin R B 1981 \textit{Phys. Rev. B} {\bf 23} 5632

\bibitem{PhysRevB.25.2185}
Halperin B I 1982 \textit{Phys. Rev. B} {\bf 25} 2185

\bibitem{RevModPhys.67.357}
Huckestein B 1995 \textit{Rev. Mod. Phys.} {\bf 67} 357

\bibitem{RevModPhys.69.315}
Sondhi S L, Girvin S M, Carini J P and Shahar D 1997 \textit{Rev. Mod. Phys.} {\bf 69} 315

\bibitem{PhysRevLett.61.1297}
Pruisken A M M 1988 \textit{Phys. Rev. Lett.} {\bf 61} 1297

\bibitem{PhysRevLett.67.883}
Koch S, Haug R J, von Klitzing K and Ploog K, 1991 \textit{Phys. Rev. Lett.} {\bf 67} 883

\bibitem{PhysRevLett.49.405}
Thouless D J, Kohmoto M, Nightingale M P and den Nijs M 1982 \textit{Phys. Rev. Lett.} {\bf 49} 405

\bibitem{Rep.Prog.Phys.64.1603}
Jeckelmann B and Jeanneret B 2001 \textit{Rep. Prog. Phys.} {\bf 64} 1603

\bibitem{Science.306.666}
Novoselov K S, Geim A K, Morozov S V, Jiang D, Zhang Y, Dubonos S V, Grigorieva I V and Firsov A A 2004 \textit{Science} {\bf 306} 666

\bibitem{RevModPhys.83.407}
Das Sarma S, Adam S, Hwang E H and Rossi E 2011 \textit{Rev. Mod. Phys.} {\bf 83} 407

\bibitem{Nature.583.375}
Stepanov P, Das I, Lu X, Fahimniya A, Watanabe K, Taniguchi T, Koppens F H L, Lischner J, Levitov L and Efetov D K 2020 \textit{Nature} {\bf 583} 375

\bibitem{Nat.Mater.6.183}
Geim A K and Novoselov K S 2007 \textit{Nat. Mater.} {\bf 6} 183

\bibitem{PhysRevB.73.235411}
Brey L and Fertig H A 2006 \textit{Phys. Rev. B} {\bf 73} 235411

\bibitem{Nature.438.197}
Novoselov K S, Geim A K, Morozov S V, Jiang D, Katsnelson M I, Grigorieva I V, Dubonos S V and Firsov A A 2005 \textit{Nature} {\bf 438} 197

\bibitem{Nature.438.201}
Zhang Y, Tan Y-W, Stormer H L and Kim P 2005 \textit{Nature} {\bf 438} 201

\bibitem{RevModPhys.80.1337}
Beenakker C W J 2008 \textit{Rev. Mod. Phys.} {\bf 80} 1337

\bibitem{PhysRevLett.97.187401}
Ferrari A C \textit{et al} 2006 \textit{Phys. Rev. Lett.} {\bf 97} 187401

\bibitem{NatPhys.2.620}
Katsnelson M I, Novoselov K S and Geim A K 2006 \textit{Nat. Phys.} {\bf 2} 620

\bibitem{PhysRevLett.95.146801}
Gusynin V P and Sharapov S G 2005 \textit{Phys. Rev. Lett.} {\bf 95} 146801

\bibitem{Science.315.1379}
Novoselov K S, Jiang Z, Zhang Y, Morozov S V, Stormer H L, Zeitler U, Maan J C, Boebinger G S, Kim P and Geim A K 2007 \textit{Science} {\bf 315} 1379

\bibitem{Science.321.385}
Lee C, Wei X, Kysar J W and Hone J 2008 \textit{Science} {\bf 321} 385

\bibitem{NatNanotechnol.3.270}
Eda G, Fanchini G and Chhowalla M 2008 \textit{Nat. Nanotechnol.} {\bf 3} 270

\bibitem{NanoLett.8.902}
Balandin A A, Ghosh S, Bao W, Calizo I, Teweldebrhan D, Miao F and Lau C N 2008 \textit{Nano Lett.} {\bf 8} 902

\bibitem{SolidStateCommun.146.351}
Bolotin K I, Sikes K J, Jiang Z, Klima M, Fudenberg G, Hone J, Kim P and Stormer H L 2008 \textit{Solid State Commun.} {\bf 146} 351

\bibitem{NatNanotechnol.3.491}
Du X, Skachko I, Barker A and Andrei E Y 2008 \textit{Nat. Nanotechnol.} {\bf 3} 491

\bibitem{Science.332.1537}
Zhu Y \textit{et al} 2011 \textit{Science} {\bf 332} 1537

\bibitem{Science.335.1326}
El-Kady M F, Strong V, Dubin S and Kaner R B 2012 \textit{Science} {\bf 335} 1326

\bibitem{Nature.575.628}
Marguerite A, Birkbeck J, Aharon-Steinberg A, Halbertal D, Bagani K, Marcus I, Myasoedov Y, Geim A K, Perello D J and Zeldov E 2019 \textit{Nature} {\bf 575} 628

\bibitem{PhysRevB.78.125409}
Ki D-K, Jeong D, Choi J-H, Lee H-J and Park K-S 2008 \textit{Phys. Rev. B} {\bf 78} 125409

\bibitem{Science.312.1191}
Berger C \textit{et al} 2006 \textit{Science} {\bf 312} 1191

\bibitem{PhysRevB.78.045322}
Levkivskyi I P and Sukhorukov E V 2008 \textit{Phys. Rev. B} {\bf 78} 045322

\bibitem{PhysRevA.41.3436}
Stern A, Aharonov Y and Imry Y 1990 \textit{Phys. Rev. A} {\bf 41} 3436

\bibitem{PhysRevB.75.081301}
Golizadeh-Mojarad R and Datta S 2007 \textit{Phys. Rev. B} {\bf 75} 081301

\bibitem{PhysRevLett.68.2512}
Meir Y and Wingreen N S 1992 \textit{Phys. Rev. Lett.} {\bf 68} 2512

\bibitem{PhysRevB.50.5528}
Jauho A-P, Wingreen N S and Meir Y 1994 \textit{Phys. Rev. B} {\bf 50} 5528

\bibitem{Datta}
Datta S 1995 \textit{Electronic Transport in Mesoscopic System} (Cambridge: Cambridge University Press) p 182

\bibitem{RevModPhys.82.1539}
Nagaosa N, Sinova J, Onoda S, MacDonald A H and Ong N P 2010 \textit{Rev. Mod. Phys.} {\bf 82} 1539

\bibitem{PhysRevLett.101.146802}
Liu C-X, Qi X-L, Dai X, Fang Z and Zhang S-C 2008 \textit{Phys. Rev. Lett.} {\bf 101} 146802

\bibitem{PhysRevLett.104.066805}
Sun Q-F and Xie X-C 2010 \textit{Phys. Rev. Lett.} {\bf 104} 066805

\bibitem{PhysRevLett.57.1761}
B\"{u}ttiker M 1986 \textit{Phys. Rev. Lett.} {\bf 57} 1761

\bibitem{PhysRevB.77.115346}
Xing Y, Sun Q-F and Wang J 2008 \textit{Phys. Rev. B} {\bf 77} 115346

\bibitem{PhysRevLett.103.036803}
Jiang H, Cheng S, Sun Q-F and Xie X-C 2009 \textit{Phys. Rev. Lett.} {\bf 103} 036803

\bibitem{PhysRevLett.63.1857}
Beenaker C W J and van Houten H 1989 \textit{Phys. Rev. Lett.} {\bf 63} 1857

\bibitem{PhysRevLett.64.216}
Beenaker C W J 1990 \textit{Phys. Rev. Lett.} {\bf 64} 216

\bibitem{PhysRevLett.61.589}
Peeters F M 1988 \textit{Phys. Rev. Lett.} {\bf 61} 589

\bibitem{PhysRevLett.95.136602}
Sheng L, Sheng D N, Ting C S and Haldane F D M 2005 \textit{Phys. Rev. Lett.} {\bf 95} 136602

\bibitem{PhysRevLett.64.220}
MacDonald A H 1990 \textit{Phys. Rev. Lett.} {\bf 64} 220

\bibitem{PhysRevB.104.115411}
Fang J-Y, Yang N-X, Yan Q, Guo A-M and Sun Q-F 2021 \textit{Phys. Rev. B} {\bf 104} 115411

\bibitem{PhysRevLett.108.166602}
Cresti A, Fogler M M, Guinea F, Castro Nteo A H and Roche S 2012 \textit{Phys. Rev. Lett.} {\bf 108} 166602

\bibitem{PhysRevB.87.235405}
Zhang Y-T, Qiao Z and Sun Q-F 2013 \textit{Phys. Rev. B} {\bf 87} 235405

\bibitem{PhysRevB.81.245417}
Chen J-C, Au Yeung T C and Sun Q-F 2010 \textit{Phys. Rev. B} {\bf 81} 245417

\bibitem{Nature.461.772}
Wang Z, Chong Y, Joannopoulos J D and Solja\v{c}i\'c M 2009 \textit{Nature} {\bf 461} 772

\bibitem{Nanolett.9.1973}
Lohmann T, von Klitzing K and Smet J H 2009 \textit{Nano Lett.} {\bf 9} 1973

\bibitem{Science.317.638}
Williams J R, DiCarlo L and Marcus C M 2007 \textit{Science} {\bf 317} 638

\bibitem{PhysRevLett.78.318}
Sheng D N and Weng Z Y 1997 \textit{Phys. Rev. Lett.} {\bf 78} 318

\bibitem{PhysRevB.73.233406}
Sheng D N, Sheng L and Weng Z Y 2006 \textit{Phys. Rev. B} {\bf 73} 233406

\bibitem{PhysRevB.59.8144}
Yang K and Bhatt R N 1999 \textit{Phys. Rev. B} {\bf 59} 8144

\bibitem{PhysRevLett.117.056802}
Qiao Z H, Han Y, Zhang L, Wang K, Deng X, Jiang H, Yang S A, Wang J and Niu Q 2016 \textit{Phys. Rev. Lett.} {\bf 117} 056802

\bibitem{PhysRevLett.76.975}
Liu D Z, Xie X C and Niu Q 1996 \textit{Phys. Rev. Lett.} {\bf 76} 975

\bibitem{RevModPhys.80.1355}
Evers F and Mirlin A D 2008 \textit{Rev. Mod. Phys.} {\bf 80} 1355

\bibitem{Science.329.61}
Yu R, Zhang W, Zhang H-J, Zhang S-C, Dai X and Fang Z 2010 \textit{Science} {\bf 329} 61

\bibitem{Science.340.167}
Chang C-Z \textit{et al} 2013 \textit{Science} {\bf 340} 167

\bibitem{NanoLett.20.7606}
Guo X, Liu Z, Liu B, Li Q and Wang Z 2020 \textit{Nano Lett.} {\bf 20} 7606

\bibitem{Science.367.895}
Deng Y, Yu Y, Shi M Z, Guo Z, Xu Z, Wang J, Chen X H and Zhang Y, 2020 \textit{Science} {\bf 367} 895

\bibitem{Science.367.900}
Serlin M, Tschirhart C L, Polshyn H, Zhang Y, Zhu J, Watanabe K, Taniguchi T, Balents L and Young A F 2020 \textit{Science} {\bf 367} 900

\bibitem{PhysRevB.82.184516}
Qi X-L, Hughes T L and Zhang S-C 2010 \textit{Phys. Rev. B} {\bf 82} 184516

\bibitem{PhysRevB.97.115452}
Zhou Y-F, Hou Z, Zhang Y-T and Sun Q-F 2018 \textit{Phys. Rev. B} {\bf 97} 115452

\end{thebibliography}
\end{document}